\documentclass{PoS}
\usepackage{textcomp,amsmath,amssymb}

\title{Summary \& Outlook: Particles and Cosmology} 
\ShortTitle{Summary \& Outlook}

\author{\speaker{Wilfried Buchm\"uller}%
       \\
       Deutsches Elektronen Synchrotron, Hamburg, Germany\\
       E-mail: \email{buchmuwi@mail.desy.de}}

\abstract{We review new results on strong and electroweak interactions, flavour physics,
cosmic rays and cosmology, which were presented at this conference, focussing on 
physics beyond the Standard Models. Special emphasis is given to the Higgs sector of
the Standard Model of Particle Physics and recent results on high-energy cosmic rays
and their implications for dark matter.}

\FullConference{European Physical Society Europhysics Conference on High Energy Physics\\
		 July 16-22, 2009\\
		 Krakow, Poland}

\begin{document}

\section{Introduction}

At this conference, the Standard Models of Particle Physics and Cosmology
have again been impressively confirmed. In the experimental talks on strong
interactions, electroweak precision tests, flavour and neutrino physics
and searches for `new physics' no significant
deviations from Standard Model predictions
have been reported. Also in Astrophysics, where unexpected results in 
high-energy cosmic rays were found, conventional astrophysical explanations
of the new data appear to be sufficient. In Cosmology, we have entered an
era of precision physics with theory lagging far behind.

Given this situation, one faces the question: What are the theoretical and
experimental hints for physics beyond the Standard Models, and what 
discoveries can we hope for at the LHC, in non-accelerator experiments,
and in astrophysical and cosmological observations? In the following
I shall summarize some results of this conference using this question as
a guideline. Particular emphasis will therefore be given to the Higgs sector
of the Standard Model, the ``topic number one'' at the LHC, and the recent 
results in high-energy cosmic rays, which caused tremendous excitement
during the past year because of the possible connection to dark matter. 

The Standard Model of Particle Physics is a relativistic quantum field theory,
a non-Abelian gauge theory with symmetry group
\begin{equation}
G_{\mathrm{SM}} = SU(3)\times SU(2)\times U(1)
\end{equation}
for the strong and electroweak interactions, respectively. Three generations
of quarks and leptons with chiral gauge interactions describe all features
of matter. The current focus is on
\begin{itemize}
\item
Precision measurements and calculations in QCD
\item
Heavy ions and nonperturbative field theory
\item
Electroweak symmetry breaking, with the key elements: top-quark,
W-boson and Higgs bosons
\item
Flavour physics and neutrinos.
\end{itemize}

The cosmological Standard Model is also based on a gauge theory, Einstein's
theory of gravity. Together with the Robertson-Walker metric this leads to
Friedmann's equations. Within current errors, the universe is known to be
spatially flat, and its expansion rate is increasing. Most remarkably,
its energy density is dominated by `dark matter' and `dark energy'. The desire 
to disentangle the nature of dark matter and dark energy, and to understand
their possible connection to particle physics is the main driving force in
observational cosmology today. 

On the theoretical frontier, string theory is the main theme, despite the fact
that after more than thirty years of research it still has not become a
falsifiable theory. Nevertheless, string theory has inspired many extensions
of the Standard Model, which will be tested at the LHC and it has
stimulated interesting models for the early universe which can be probed by
cosmological observations. String theory goes beyond field theory by replacing
point-interactions of particles by nonlocal interactions of strings. In this
way it has also become a valuable tool to analyze strongly interacting
systems of particles at high energies and high densities.

\section{Strong Interactions}

\subsection{QCD at colliders}

Quantum chromodynamics is the prototype of a non-Abelian gauge theory. 
To improve our quantitative understanding of this theory has remained a
theoretical challenge for more than three decades. In recent years important
topics have been the determination of the scale-dependent strong coupling 
$\alpha_s(Q^2)$, higher-order calculations of matrix elements, the analysis
of multi-leg final states and soft processes including underlying events and
diffraction \cite{schleper}. 

\begin{figure}[t]
\begin{center}
\includegraphics[width=6.5cm]{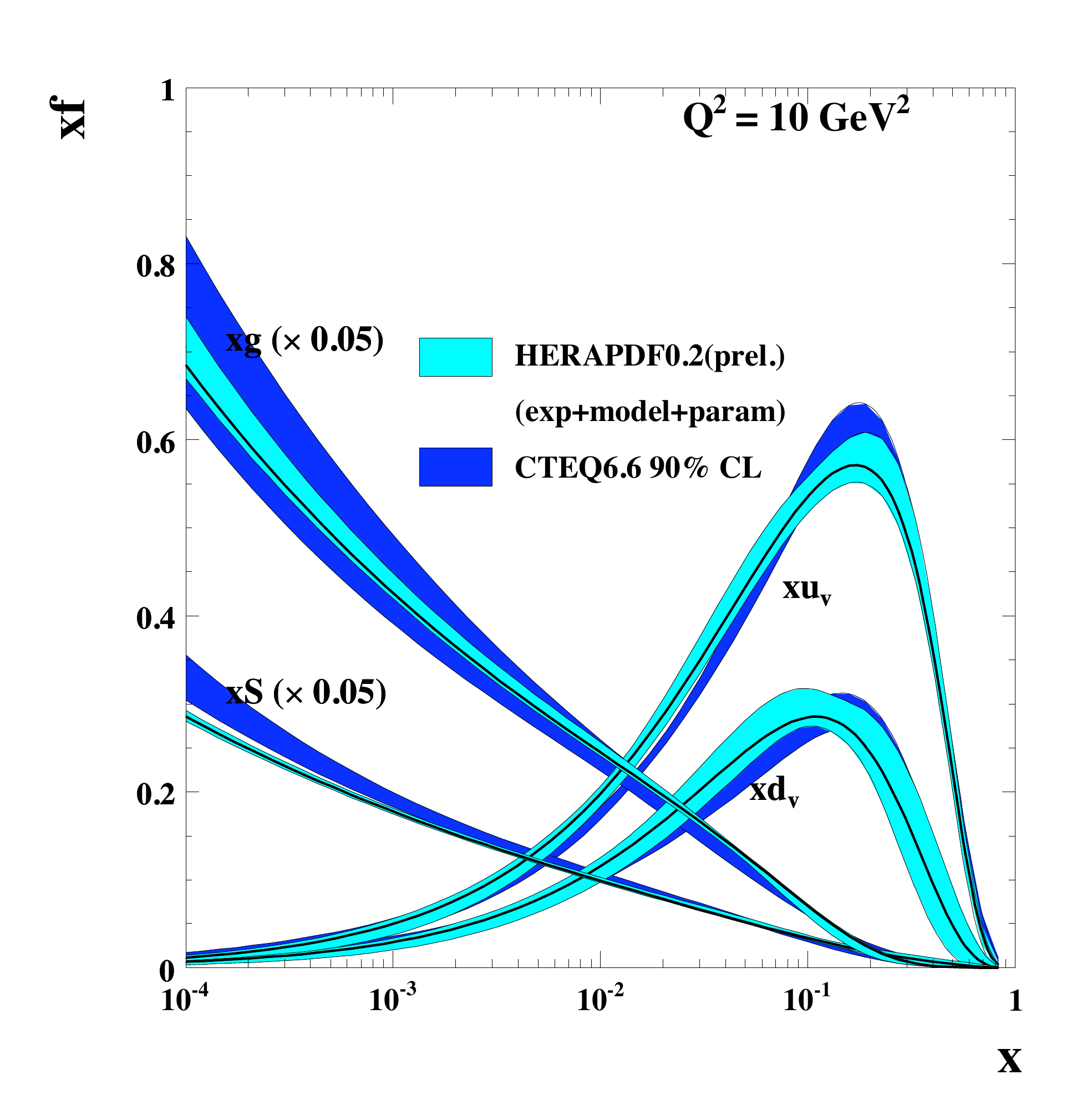}
\hspace*{1cm}
\includegraphics[width=6.5cm]{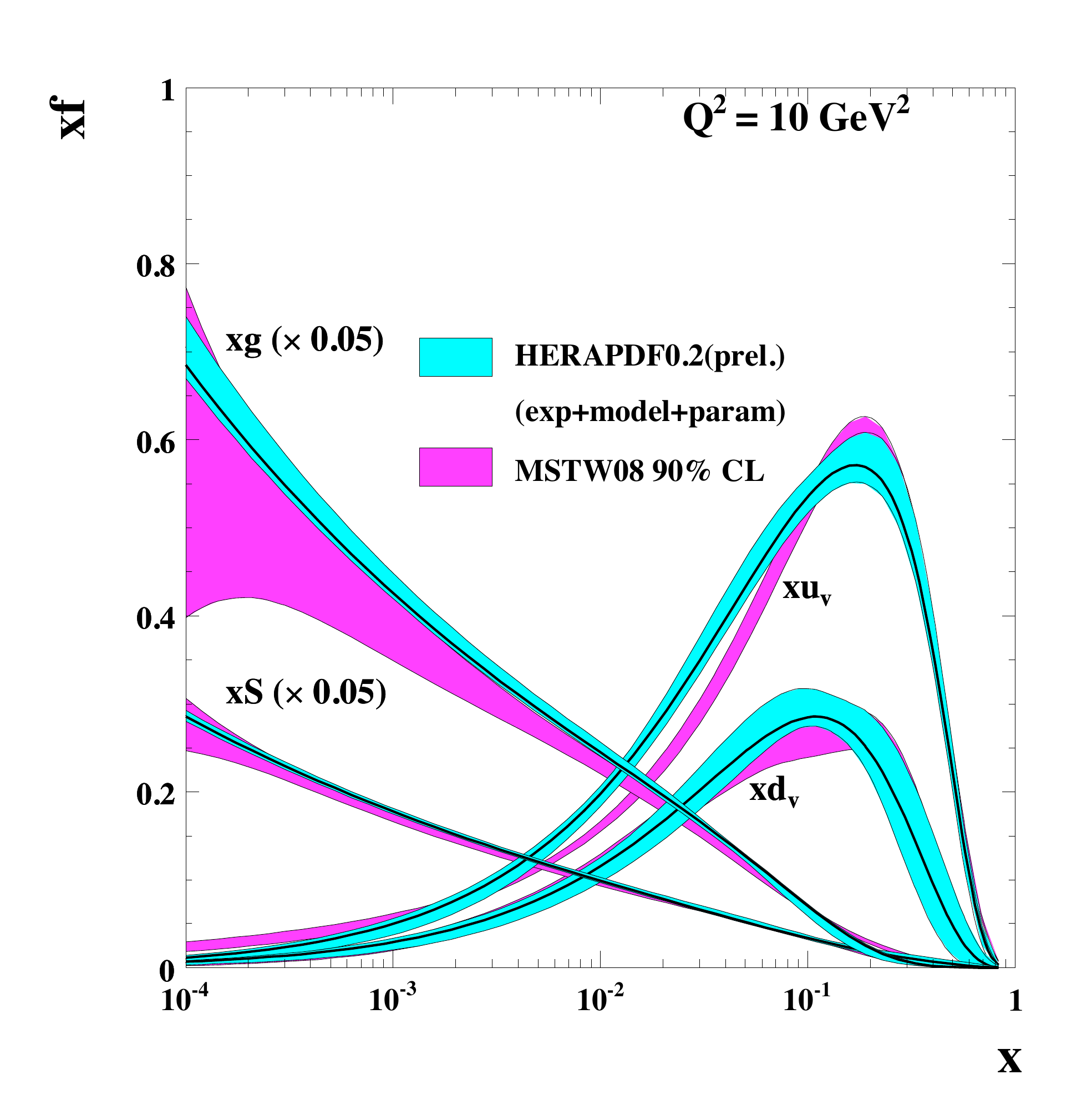}
\end{center}
\caption{Quark and gluon distribution functions from a combined analysis of
the H1 and ZEUS collaborations compared with distribution functions obtained
by CTEQ (left) and MSTW (right). From \cite{schleper}.
\label{fig:PDFs}
}
\end{figure}

Understanding QCD is also a prerequisite for electroweak 
precision tests and physics
beyond the Standard Model. The search for the Higgs boson, for instance,
requires the knowledge of the gluon distribution function, at low Bjorken-$x$
for light Higgs bosons and at large Bjorken-$x$ for large Higgs masses. 
Recently, a combined analysis of the deep-inelastic scattering data of the H1 
and ZEUS collaborations at HERA has led to significantly more precise quark 
and gluon distribution functions in the whole $x$-range. The new HERA-PDF's are
compared with previous determinations of parton distribution functions
by the CTEQ and MSTW collaborations
in Fig.~\ref{fig:PDFs}. 
  
\begin{figure}[t]
\begin{center}
\includegraphics[height=6cm]{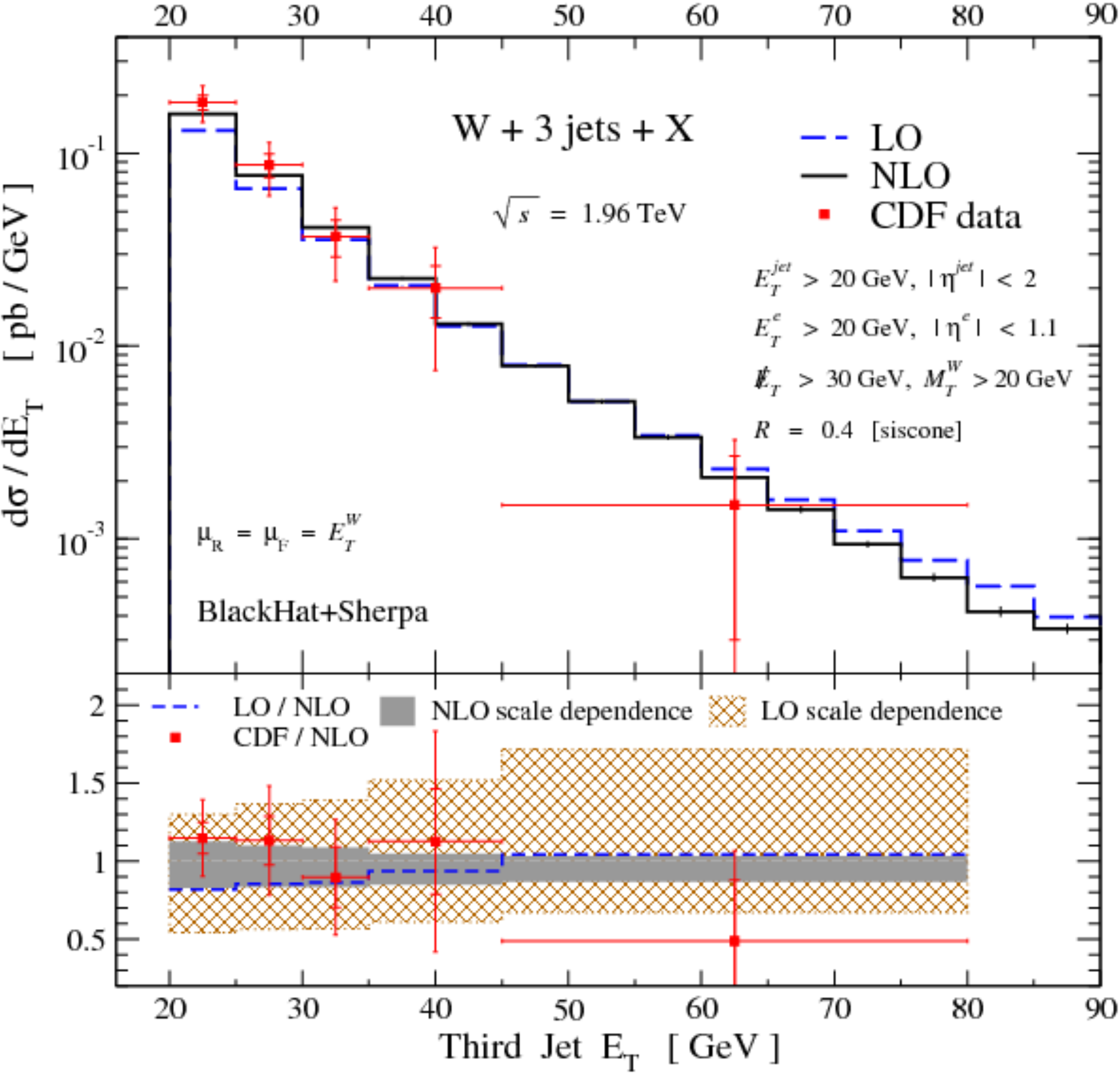}
\hspace*{1cm}
\includegraphics[height=6.3cm]{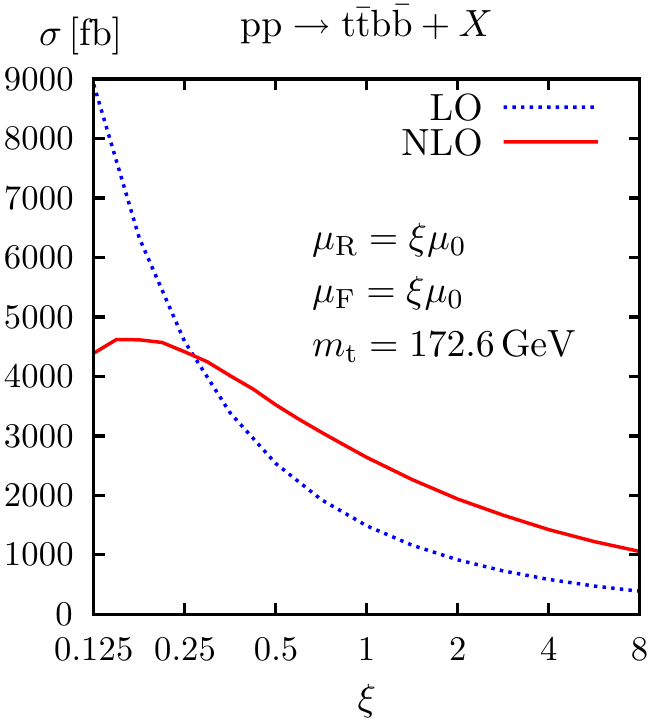}
\end{center}
\caption{Left: The measured inclusive W+3jet production cross section for
$p\bar{p}$ collisions at the Tevatron as function of the Third Jet $E_T$;
from \cite{schleper}. Right: Scale dependence of LO and NLO cross sections
for the process $pp \rightarrow t\bar{t}b\bar{b} + X$ at the LHC. 
From \cite{anastasiou}.
\label{fig:NLO}
}
\end{figure}

Impressive progress has been made in the development of new techniques for
multi-leg next-to-leading (NLO) calculations \cite{anastasiou}. As a
result, the full NLO calculation for the inclusive W+3jet production cross
section in hadron-hadron collisions became possible. In Fig.~\ref{fig:NLO}
the LO and NLO predictions are compared with CDF data; the scale dependence 
is significantly reduced. Another important process, especially as
background for Higgs search, is
$pp \rightarrow t\bar{t}b\bar{b} + X$ for which a full NLO calculation
has also been performed. As expected, the scale dependence is reduced
(see Fig.~\ref{fig:NLO}). 
One may worry, however, that the `correction' compared to the LO calculation
is $\mathcal{O}(100)\%$! Most remarkable is also the progress in calculating
multi-leg amplitudes. Using conventional as well as string theory techniques
it has become possible to compute scattering amplitudes involving up to 22 
gluons \cite{anastasiou}!

\subsection{Quark-gluon plasma and AdS/CFT correspondence}

During the past years dense hadronic matter has become another frontier of 
QCD due to new results from RHIC and novel theoretical developments
\cite{wiedemann}. An interesting collective phenomenon is the `elliptic
flow' of particles produced in heavy ion collisions. From the size and
$p_T$-dependence of the elliptic flow one can determine the shear viscosity
$\eta$ which appears as parameter in hydrodynamic simulations. The small
measured value of $\eta$ caused considerable excitement among theorists
since it could be understood in the context of a strongly coupled 
$N=4$ supersymmetric Yang-Mills (SYM) theory.

\begin{figure}[b]
\begin{center}
\includegraphics[width=9cm]{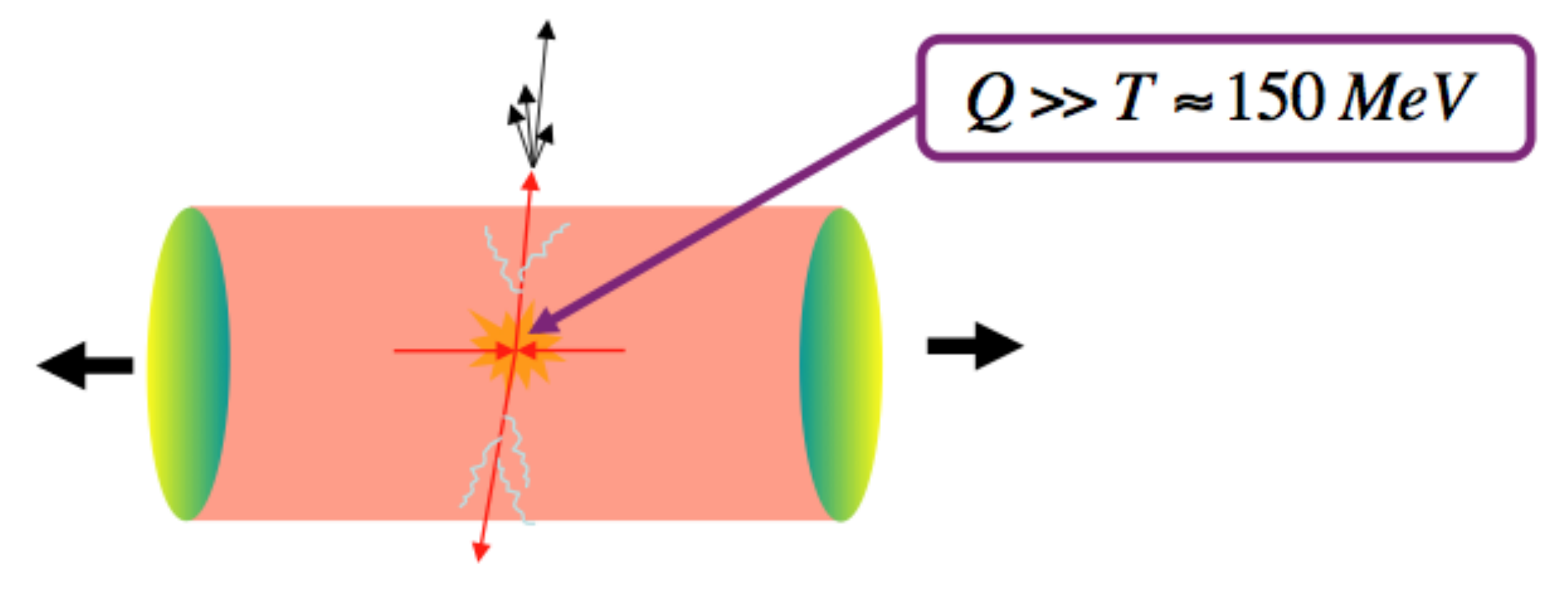}
\end{center}
\caption{Monojet event in a heavy ion collision. From \cite{wiedemann}.}
\label{fig:mono}
\end{figure}

Another intriguing phenomenon are monojets, originally conjectured by Bjorken 
for proton-proton collisions. In heavy ion collisions their appearance is
expected due to the radiative energy loss  in the medium (see
Fig~\ref{fig:mono}), which can be asymmetric for partons after a hard 
scattering. The monojet phenomenon has already been observed at RHIC
and will be studied in detail at LHC.
 
\begin{figure}[t]
\begin{center}
\includegraphics[width=2.5cm]{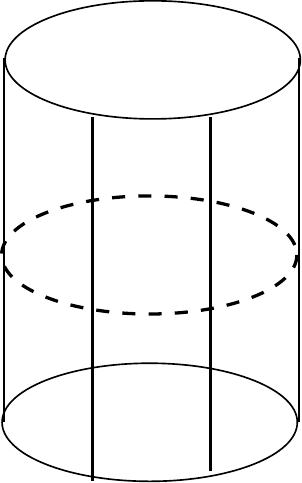}
\end{center}
\caption{The leading L\"uscher graph contributing to the Konishi operator at 
four loops. The dashed line represents all asymptotic states of the theory 
while the two vertical lines correspond to the two particles forming the 
Konishi state in the two dimensional string worldsheet QFT. From \cite{janik}.}
\label{fig:luscher}
\end{figure}

On the theoretical side, significant progress has been made towards `solving
$N=4$ SYM theory' \cite{janik}. Here `solving' means the determination of
the anomalous dimensions of all operators for any value of the gauge coupling,
so that one can extrapolate the theory from the perturbative weak-coupling
regime to the nonperturbative strong coupling regime. The anomalous dimensions
can be calculated in usual perturbation theory as well as, via the AdS/CFT 
correspondence, by means of string theory in the spacetime background
$AdS_5\times S^5$, i.e., by considering a particular two-dimensional field
theory on a finite cylinder (see Fig.~\ref{fig:luscher}). 
 
As an example, consider the Konishi operator $\mathrm{tr}\Phi_i^2$, where 
$\Phi_i$ 
are the adjoint scalars of $N=4$ SYM. String theory yields at order $g^8$,
\begin{equation}
\label{sum}
\Delta^{(4-loop)}_{wrapping} = 
\sum_{Q=1}^{\infty} \left\{ -\frac{num(Q)}{\left(9 Q^4-3
   Q^2+1\right)^4 \left(27 Q^6-27 Q^4+36 Q^2+16\right)}
+\frac{864}{Q^3}-\frac{1440}{Q^5} \right\}\ ,
\end{equation}
with the numerator 
\begin{align}
num(Q) =& 7776 Q (19683 Q^{18}-78732 Q^{16}+150903
   Q^{14}-134865 Q^{12}+ \nonumber\\
&+1458 Q^{10}+48357 Q^8-13311
   Q^6-1053 Q^4+369 Q^2-10)\ .
\end{align}
The sum (\ref{sum}) can be carried out with the result  
\begin{equation}
\Delta^{(4-loop)}_{wrapping} = (324+864\zeta(3)-1440 \zeta(5))g^8\ ,
\end{equation}
which exactly agrees with a direct perturbative computation at four-loop order
(around 131015 Feynman graphs). The recent string calculation to order $g^{10}$
still remains to be checked by a five-loop perturbative calculation.

The string calculations give the impression that there is some structure in
the perturbative expansion of gauge theories which has not been understood
so far. In this way, $N=4$ SYM theory may become the `harmonic oscillator
of four-dimensional gauge theories'.

\section{The Higgs sector}

The central theme of physics at the LHC is the Higgs sector \cite{grojean}
of the Standard Model. The weak and electromagnetic interactions are described 
by a spontaneously broken gauge theory. The Goldstone bosons of the symmetry
breaking
\begin{equation}\label{sym1}
SU(2)_L\times U(1)_Y  \rightarrow U(1)_\textrm{em}  
\end{equation}
give mass to the W- and Z-bosons via the Higgs mechanism. In the Standard Model
the electroweak symmetry is broken in the simplest possible way, by the vacuum 
expectation value of a single $SU(2)_L$ doublet, corresponding to the symmetry
breaking $SU(2)_L\times SU(2)_R \rightarrow SU(2)_{L+R}$ which contains the 
Goldstone bosons of (\ref{sym1}). 

The unequivocal prediction of the Standard Model is the existence of a new
elementary particle, the Higgs boson. During the past two decades this theory
has been impressively confirmed in many ways. Electroweak precision tests
favour a light Higgs boson \cite{conway}, 
$m_H \simeq 87^{+35}_{-26}\ \textrm{GeV}$. 
For Higgs masses in the range from 130~GeV to 180~GeV, the
Standard Model can be consistently extrapolated from the electroweak scale
$\Lambda_\textrm{EW} \sim 100~\textrm{GeV}$ to the grand unification (GUT)
scale $\Lambda_\textrm{GUT} \sim 10^{16}~\textrm{GeV}$, avoiding the potential
problems of vacuum instability and the Landau pole for the Higgs self-coupling.

The unsuccessful search for the Higgs boson
at LEP led to the lower bound on the Higgs mass $114\ \textrm{GeV} < m_H$,
and the search at the Fermilab Tevatron excludes the the mass range 
$163 < m_H < 166\ \textrm{GeV}$ at 95\% C.L.
(see Fig.~{\ref{fig:higgstevatron}). Together with the assumption of grand
unification, and the implied upper bound of $180\ \textrm{GeV}$ on 
the Higgs mass, this further supports the existence of a light Higgs boson.

\begin{figure}[t]
\begin{center}
\includegraphics[width=0.6\textwidth,clip,angle=0]{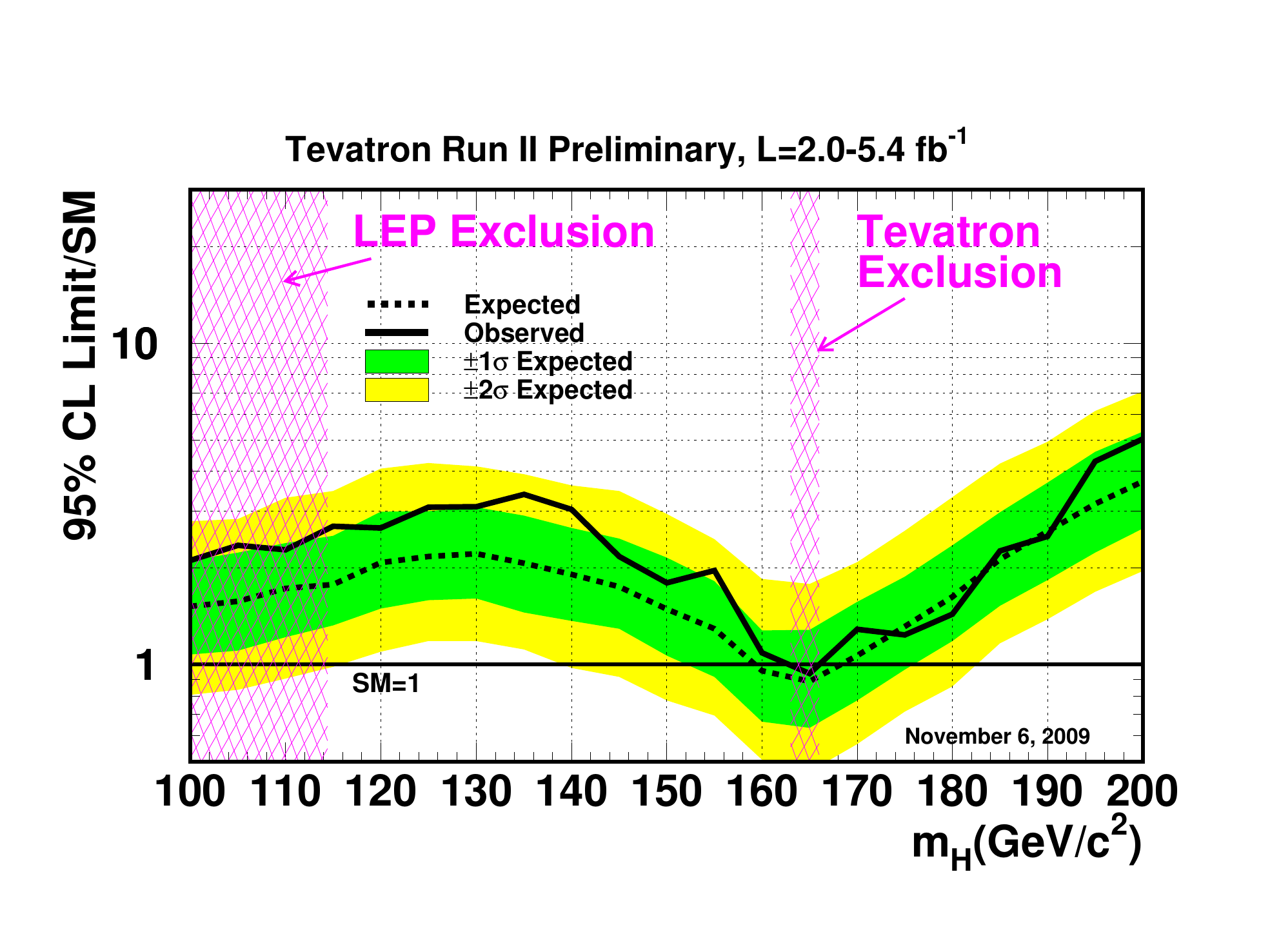}
%\hspace*{-0.5cm}
\caption{ \label{fig:higgstevatron} 
Observed and expected 95\% C.L. on the ratios to the SM cross sections, as 
functions of the Higgs boson mass; combined CDF and D0 analysis.
From \cite{conway}.}
\end{center}
\end{figure}

Supersymmetric extensions of the Standard Model are particularly well 
motivated. They stabilize the hierarchy between the electroweak scale and the
GUT scale, and in the minimal case with two $SU(2)_L$ doublets, 
the MSSM, the strong and electroweak gauge couplings unify with surprizing 
accuracy at $\Lambda_\textrm{GUT} \sim 10^{16}~\textrm{GeV}$. In addition, 
the lightest supersymmetric particle (LSP), neutralino or gravitino, is a
natural dark matter candidate. As a consequence, the search for superparticles
dominates `New Physics' searches at the Tevatron and the LHC \cite{buescher}.

On the other hand, supersymmetric versions of the Standard Model require some
`fine-tuning' of parameters. In particular in the MSSM, where the Higgs 
self-coupling is given by the gauge couplings, one obtains at tree level an 
upper bound on the lightest CP-even Higgs scalar,
\begin{equation}
m_h \leq m_Z\ .
\end{equation} 
One-loop radiative corrections can lift the Higgs mass above the LEP bound 
provided the scalar top is heavier than 1~TeV. Consistency with the 
$\rho$-parameter then requires an adjustement of different parameters at the level
of 1\%, which is sometimes considered to be `unnatural'
\footnote{Note, that in the non-supersymmetric Standard Model the small
value of the CP-violating parameter $\epsilon'$ is also due to fine-tuned 
cancellations between unrelated contributions.}. This fine-tuning can be
avoided in models with more fields such as the next-to-minimal supersymmetric
Standard Model (NMSSM) or `little Higgs' models, where the Higgs fields 
appear as pseudo-Goldstone bosons of a global symmetry containing
$SU(2)_L\times SU(2)_R \rightarrow SU(2)_{L+R}$. 

So far no Higgs-like boson has been found and we do not know what the 
origin of electroweak symmetry breaking is. Theorists have been rather 
inventive and the considered possibilities range from weakly coupled 
elementary Higgs bosons with or without supersymmetry via composite Higgs 
bosons and technicolour to the extreme case of large extra dimensions with
no Higgs boson. The corresponding Higgs scenarios come with colourful
names such as \cite{grojean}: 
buried, charming, composite, fat, fermiophobic, gauge, gaugephobic, 
holographic, intermediate, invisible, leptophilic, little, littlest, lone, 
phantom, portal, private,  slim, simplest, strangephilic, twin, un-, unusual, 
\ldots .
The various possibilities will hopefully soon be reduced by LHC data.

\subsection{Weak versus strong electroweak symmetry breaking}

To unravel the nature of electroweak symmetry breaking, it is not sufficient 
to find a `Higgs-like' resonance and to measure mass and spin. Of crucial
importance is also the study of longitudinally polarized W-bosons at large
center-of-mass energies, $s \gg m_W^2$, a notoriously difficult measurment. 
The gauge boson self-interactions lead to a $WW$ scattering amplitude which 
rises with energy, 
\begin{equation}
\mathcal{A} (W_L^a W_L^b \to W_L^c W_L^d) = 
\mathcal{A}(s) \delta^{ab}\delta^{cd} 
+ \mathcal{A}(t) \delta^{ac}\delta^{bd} 
+\mathcal{A}(u) \delta^{ad}\delta^{bc}\ , \qquad  
\mathcal{A}(s)= i\frac{s}{v^2},
\end{equation}
and violates perturbative unitarity at $\sqrt{s} = 1-3\ \mathrm{TeV}$.
A scalar field $h$, which couples to longitudinal $W$'s with strength $\alpha$
relative to the SM Higgs coupling, yields the additional scattering amplitude 
\begin{equation}
\mathcal{A}_\textrm{\tiny scalar}(s) = 
- i\frac{\alpha^2}{v^2(s-m_h^2)}\ .
\end{equation}
As expected, the leading term of the total scattering amplitude,  
\begin{equation}
\mathcal{A}_\textrm{\tiny tot}(s) = 
- i\frac{(\alpha^2-1)s^2 + m_h^2 s}{v^2 (s-m_h^2)}\ ,
\end{equation}
vanishes for $\alpha^2=1$, which corresponds to the SM Higgs, and unitarity 
is restored. It is important to realize, however, that the exchange of a
scalar may only partially unitarize the $WW$ scattering amplitude. This 
happens in composite Higgs models where the Higgs mass can be light compared 
to the compositeness scale $f>v$. Restoration of unitarity is then postponed
to energies $\sqrt{s} \sim 4\pi f > m_h$, where additional degrees of freedom 
become visible, which are related to the strong interactions forming the 
composite Higgs boson.  

\begin{figure}[t]
\begin{center}
\includegraphics[width=0.45\textwidth,clip,angle=0]{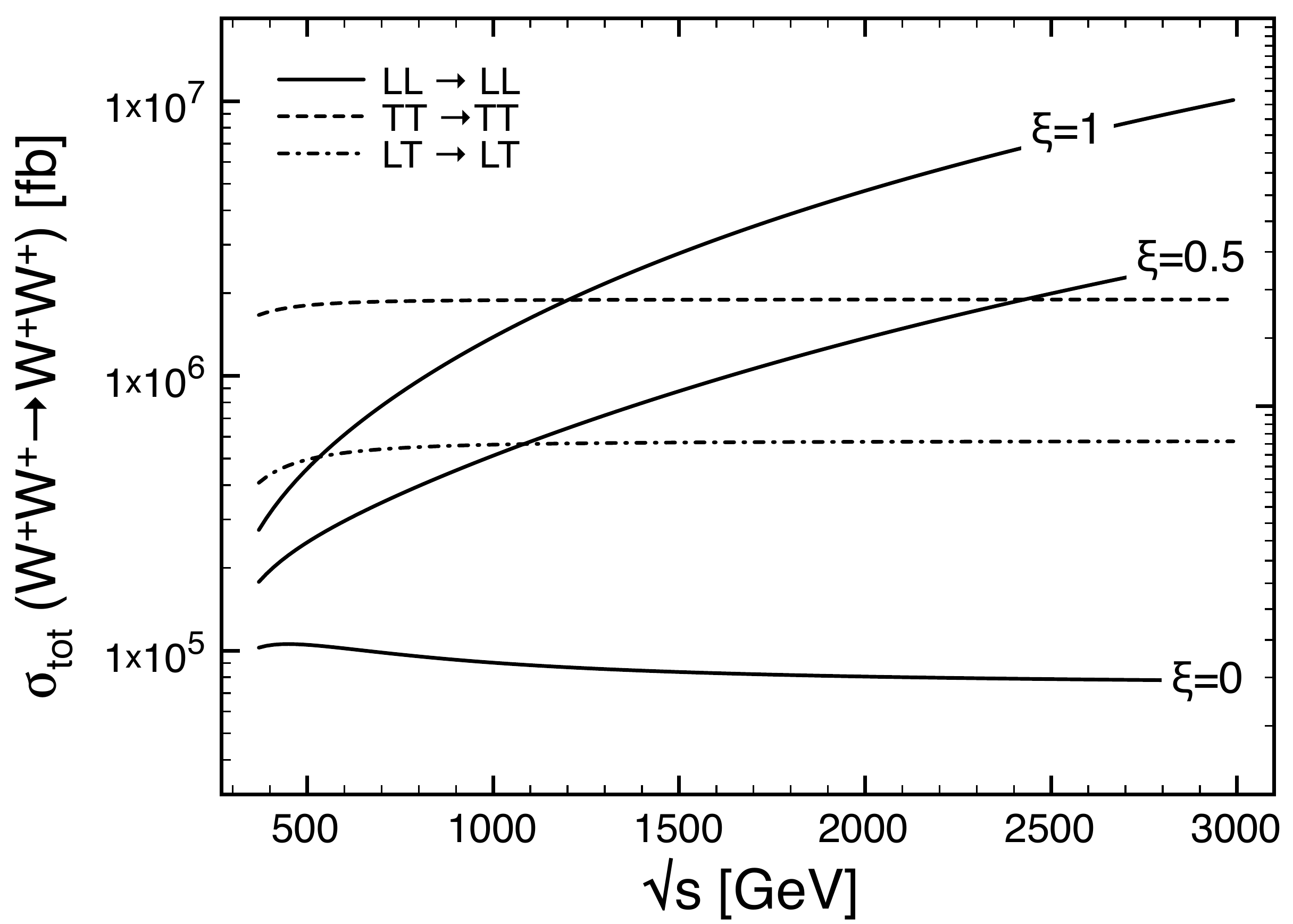}
\hspace{.2cm}
\includegraphics[width=0.45\textwidth,clip,angle=0]{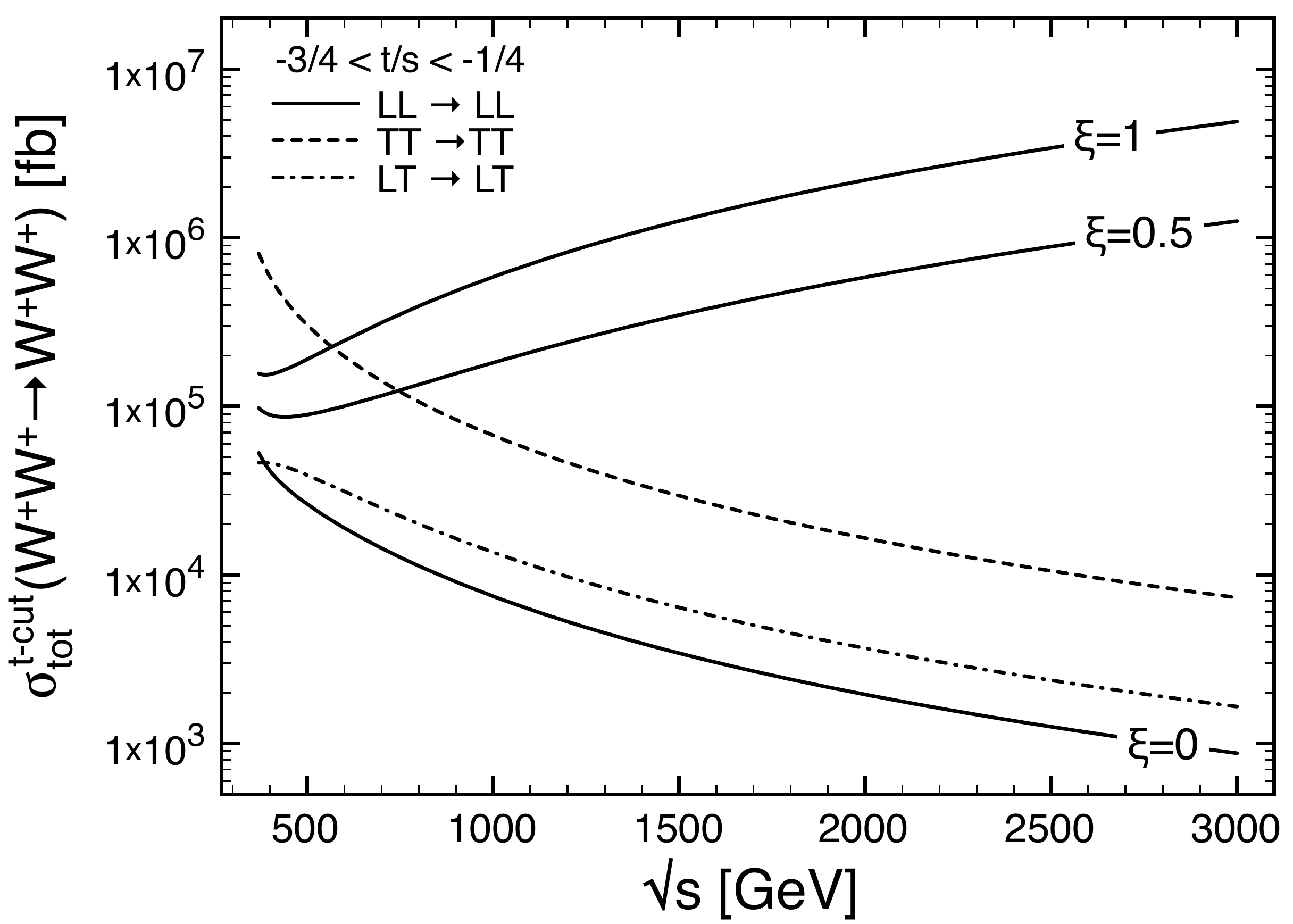}
\caption{\label{fig:WpWpTOWpWp}
 $W^+W^+\to W^+W^+$ partonic cross section as a function of the center-of-mass 
energy for $m_h = 180$~GeV for the SM ($\xi=0$) and for composite Higgs models 
($\xi=v^2/f^2 \not =0$).
On the left, the inclusive cross section is shown with a cut on $t$ and $u$ 
of order $m_W^2$; the plot on the right displays the hard cross section with 
the cut $-0.75 < t/s < -0.25$. From \cite{grojean}.  
}\label{fig:unitarity}
\end{center}
\end{figure}

Signatures of composite Higgs models can be systematically studied by adding
higher-dimensional operators to the Standard Model Lagrangian \cite{grojean},  
\begin{eqnarray}
&&\mathcal{L}_{\tiny comp} = 
\frac{c_H}{2f^2} \left( \partial_\mu \left( H^\dagger H \right) \right)^2
+ \frac{c_T}{2f^2}  \left(   H^\dagger{\overleftrightarrow D}_\mu H\right)^2 
- \frac{c_6\lambda}{f^2}\left( H^\dagger H \right)^3 
+ \left( \frac{c_yy_f}{f^2}H^\dagger H  {\bar f}_L Hf_R +{\rm h.c.}\right) \nonumber \\ 
&&
+\frac{ic_Wg}{2m_\rho^2}\left( H^\dagger  \sigma^i \overleftrightarrow {D^\mu} H \right )( D^\nu  W_{\mu \nu})^i
+\frac{ic_Bg'}{2m_\rho^2}\left( H^\dagger  \overleftrightarrow {D^\mu} H \right )( \partial^\nu  B_{\mu \nu})  +\ldots \ ;
\label{eq:comp}
\end{eqnarray}
here $g, g', \lambda$ and $f_{L,R}$ are the electroweak gauge couplings, the 
quartic Higgs coupling and the Yukawa coupling of the fermions $f_{L,R}$,
respectively; $m_{\rho} \simeq 4\pi f$, and the coefficients, 
$c_H, c_T \ldots$ are expected to be of order one. The effective Lagrangian
(\ref{eq:comp}) describes departures from the Standard Model to leading order
in $\xi = v^2/f^2$.

The measurement of a rising cross section for longitudinal W-bosons at the
LHC is a challenging task. In Fig.~\ref{fig:unitarity} the predicted rise with
energy is shown for $m_h = 180$~GeV and two values of $\xi = v^2/f^2$.
The discovery of a `Higgs boson' at the LHC, with no other signs of new
physics, would still allow a rather low scale of compositeness, 
$\xi \simeq 1$. 

\begin{figure}[htbp]
\begin{center}
\includegraphics[width=.75\textwidth]{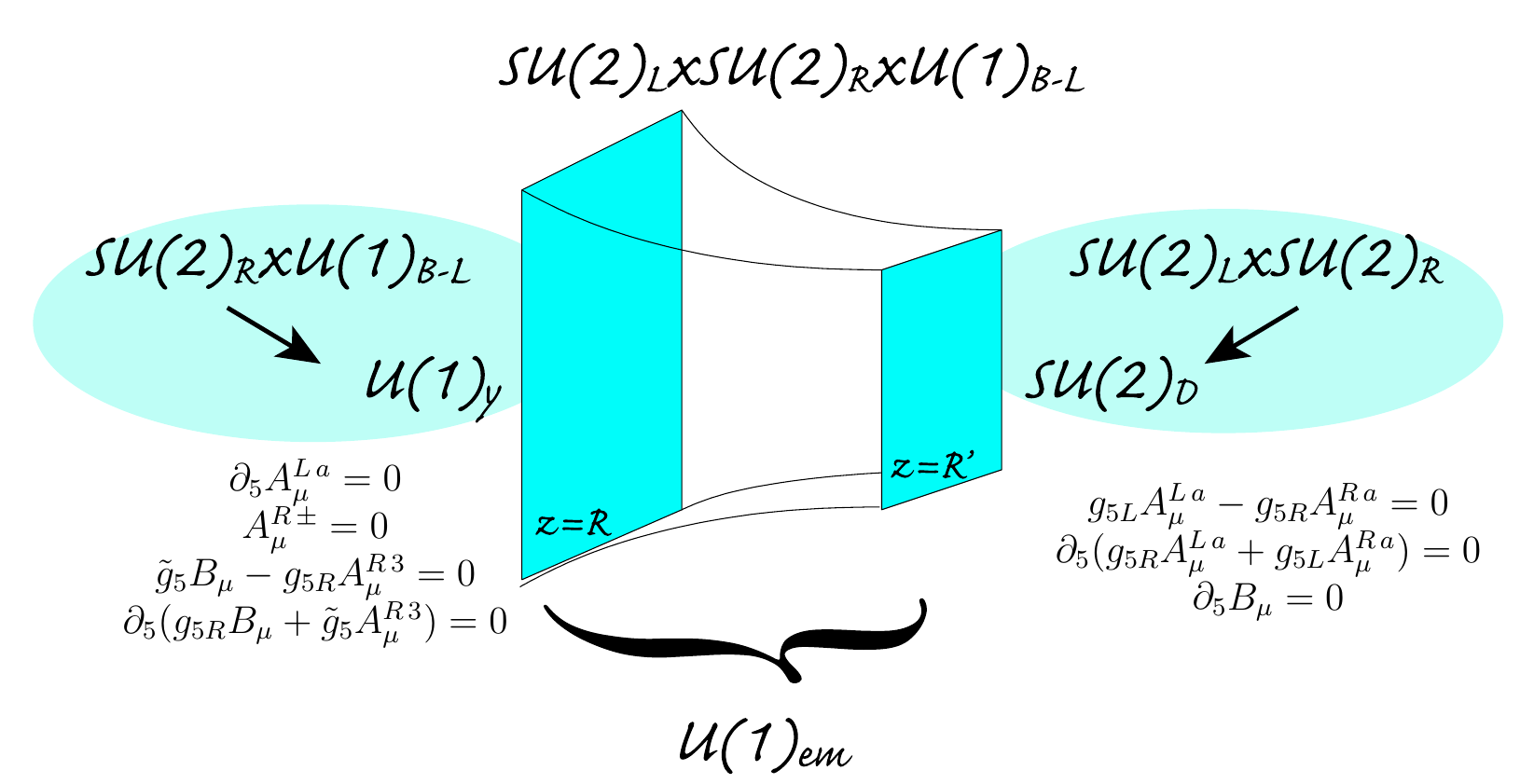}
\caption[]{The symmetry-breaking structure of the warped Higgsless
model of Csaki et al. The model considers a 5D gauge theory in a fixed gravitational anti-de-Sitter (AdS) background.
The UV~brane (sometimes called the Planck brane) is located at $z=R$ and the IR~brane (also called the TeV brane) is located at $z=R'$. $R$ is the AdS curvature scale. In conformal coordinates, the AdS metric is given by
$
ds^2=  \left( R/z \right)^2   \Big( \eta_{\mu \nu} dx^\mu dx^\nu - dz^2 \Big).
$ From \cite{grojean}.
}
\label{fig:higgsless}
\end{center}
\end{figure}

\subsection{Higgsless models}

Despite all electroweak precision tests, it still is conceivable that the 
electroweak gauge symmetry is broken without Higgs mechanism and that no Higgs 
boson exists. However, in this extreme case other new particles are predicted, 
which unitarize the $WW$ scattering amplitude.

An interesting example of this kind are higher-dimensional theories with size
of the electroweak scale, $r_{\tiny higgsless} \sim 1/v = 
\mathcal{O}(10^{-16}\ \mathrm{cm}$) (see Fig.~\ref{fig:higgsless}). The $W$- and
$Z$-bosons are now interpreted as Kaluza-Klein modes whose mass is due to
their transverse momentum in the extra dimensions,  
\begin{equation}
E^2 = \vec{p}^2_3 + p^2_\perp = \vec{p}^2_3 + m_W^2 \ ,
\end{equation}
where $\vec{p}$ is the ordinary 3-momentum. Naively, one expects a strong
rise of the $WW$ scattering amplitude with energy,
\begin{equation}
\mathcal{A} = \mathcal{A}^{(4)} \left( \frac{\sqrt{s}}{v}\right)^4 
+ \mathcal{A}^{(2)} \left( \frac{\sqrt{s}}{v}\right)^2 + \mathcal{A}^{(0)} + \ldots 
\end{equation}
However, inclusion of all Kaluza-Klein modes leads to 
$\mathcal{A}^{(4)}=\mathcal{A}^{(2)}=0$, which is a consequence of the 
relations between couplings and masses enforced by the higher-dimensional
gauge theory. Since the extra dimensions have electroweak size, higgsless
models predict $W'$ and $Z'$ vector bosons below 1~TeV with sizable
couplings to the standard $W$ and $Z$ vector bosons.

\subsection{Top-Higgs system}

\begin{figure}[t]
\begin{center}
\includegraphics[height=7cm]{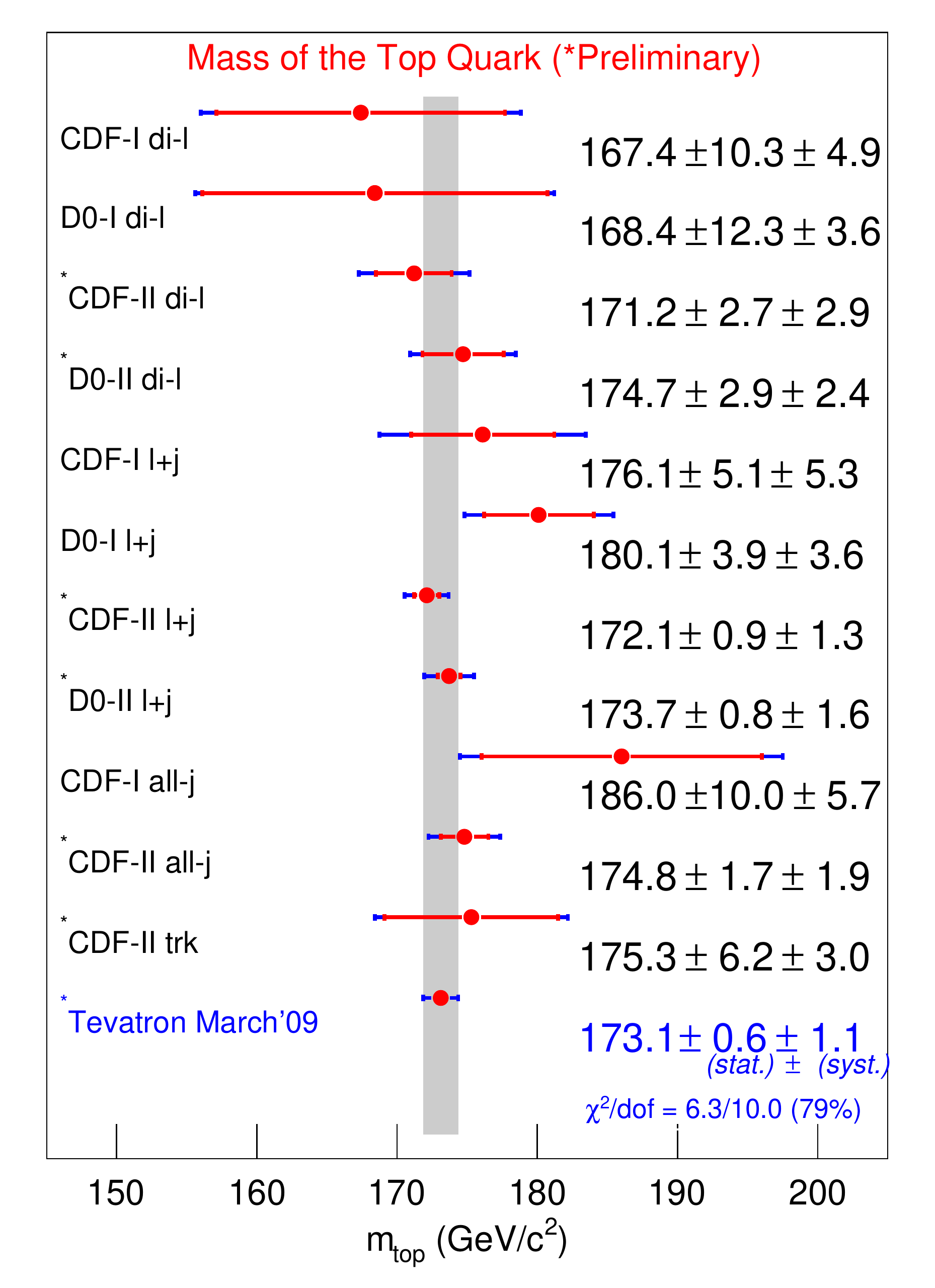}
\hspace*{1cm}
\includegraphics[height=8cm]{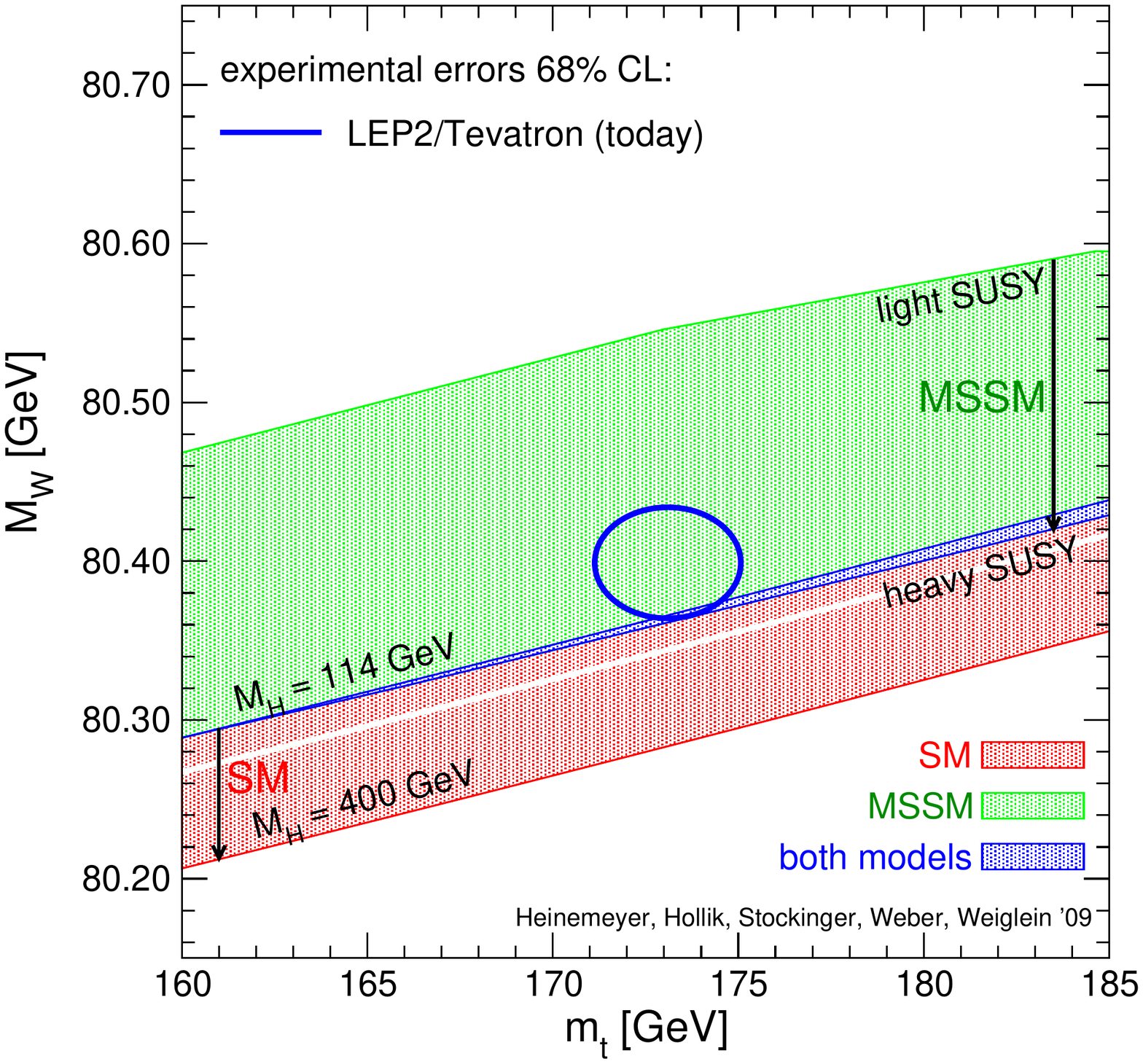}
\caption{ \label{fig:tophiggs}
Left: Top-quark mass measurements of CDF and D0. Right: Predicted dependence
of the W-mass on the top mass in the SM and the MSSM. From 
\cite{schwanenberger}. 
}
\end{center}
\end{figure}

In the Standard Model the top-quark \cite{schwanenberger} plays a special role 
because of its large
Yukawa coupling. In some supersymmetric extensions, the top Yukawa coupling
even triggers electroweak symmetry breaking. It is very remarkable that
the top-quark mass is now known with an accuracy comparable to its width
(see Fig.~\ref{fig:tophiggs}),
\begin{equation}
m_{\textrm{top}} = 173.1 \pm 1.3\ \mathrm{GeV}\ .
\end{equation}
The meaning of a top-quark mass given with this precision is a subtle
theoretical issue. To further improve this precision would be very interesting
for several reasons. First of all, it is a challenge for the present 
theoretical understanding of QCD processes to relate the measured `top-quark 
mass' to parameters of the Standard Model Lagrangian. Moreover, since the
top-Higgs system plays a special role in many extensions of the Standard
Model, one may hope to discover some departure form Standard Model predictions.

In the right panel of Fig.~\ref{fig:tophiggs} the predicted dependence of the
$W$-mass \cite{hays} on the top mass is compared for the SM and the MSSM. 
It is intriguing
that, at the 68\% C.L., the supersymmetric extension of the Standard Model is
favoured, but clearly increased precision is needed \cite{hays}.

\section{Flavour Physics}

The remarkable success of the CKM description of flavour violation and in
particular CP violation is demonstrated by the so-called Unitarity Triangle
fit shown in Fig.~\ref{fig:CKM}. A large data set on quark mixing angles
and CP-asymmetry parameters are consistent within theoretical and experimental
uncertainties \cite{bevan}. So far no deviation from the Standard Model 
has been detected.
Via a naive operator analysis, one obtains from electroweak precision tests 
and data on flavour changing neutral currents (FCNC) the lower bounds on
`new physics' \cite{buras}:
\begin{equation}
\Lambda^{\mathrm{EW}}_{\mathrm{NP}} > 5~\mathrm{TeV}\ ,\quad
\Lambda^{\mathrm{FCNC}}_{\mathrm{NP}} > 1000~\mathrm{TeV}\ .
\end{equation}
Hence, it may very well be that no departures from the Standard Model will
be found at the LHC and other currently planned accelerators.

\begin{figure}[t]
\begin{center}
\includegraphics[width=0.45\textwidth,clip,angle=0]{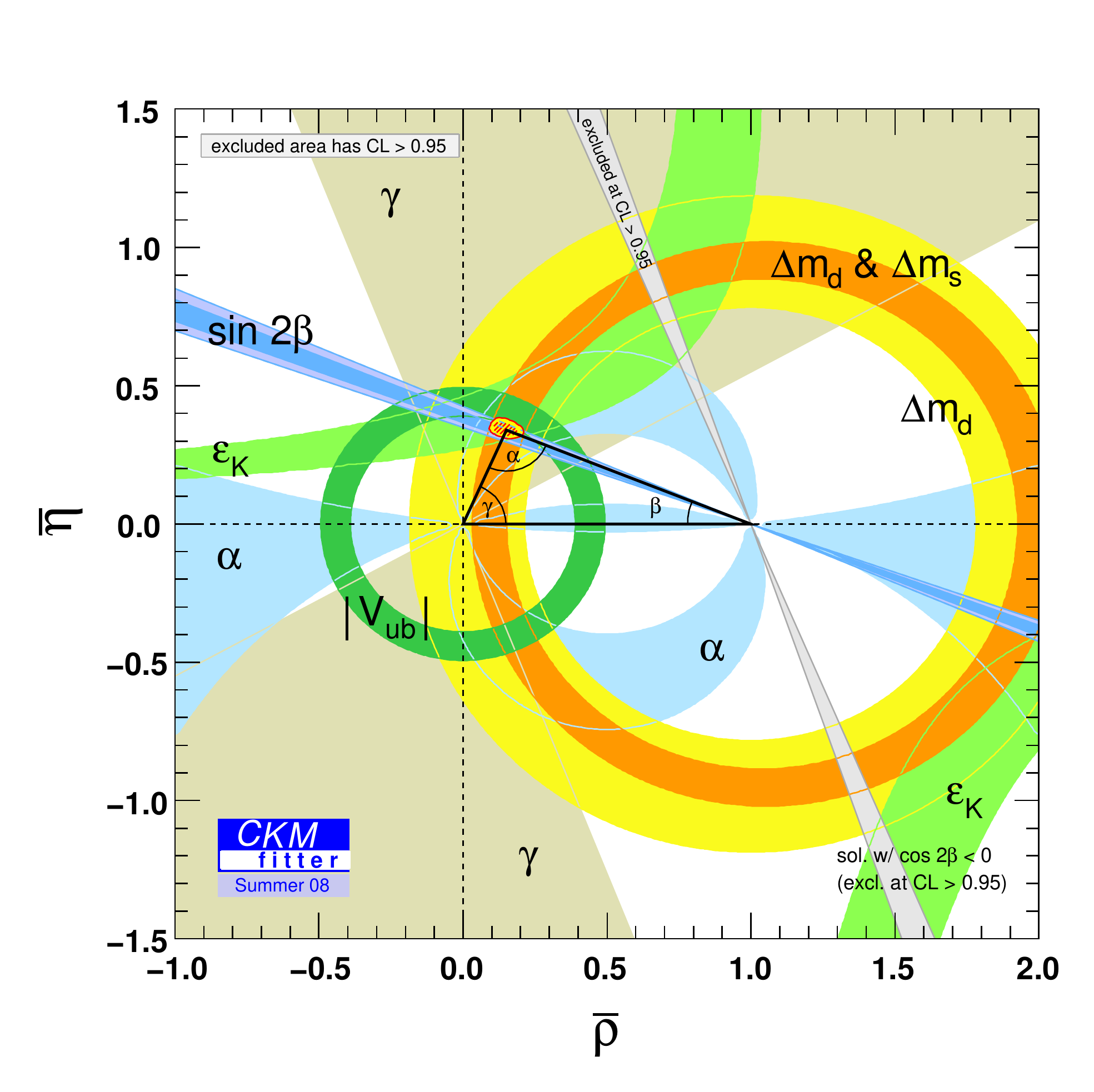}
%\hspace*{-0.5cm}
\caption{ \label{fig:CKM} 
Unitarity triangle fit by the CKMfitter collaboration in 2009.
From \cite{buras}.}
\end{center}
\end{figure}

On the other hand, as we have already seen in our discussion of the Higgs
sector, it is also conceivable that dramatic departures from the Standard
Model will be discovered at the LHC. In this case new physics in FCNC 
processes is also expected at TeV energies. This is the case in supersymmetric
extensions of the Standard Model, the `Littlest Higgs' model with T-parity
or Rundall-Sundrum models, as discussed in detail in \cite{buras}.

\begin{figure}[b]
\begin{center}
\includegraphics[width=0.4\textwidth,clip,angle=0]{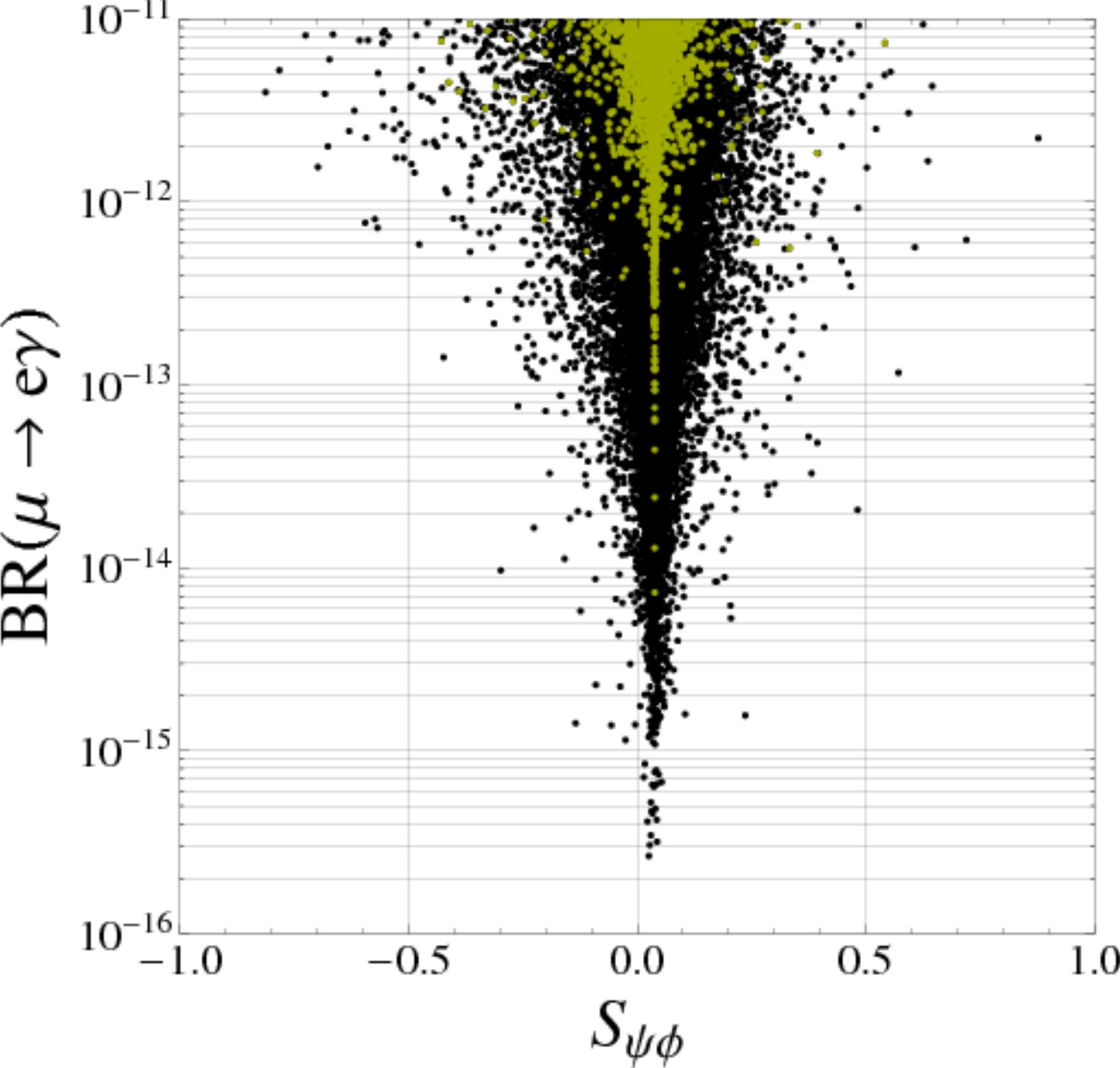}
\hspace{1cm}
\includegraphics[width=0.4\textwidth,clip,angle=0]{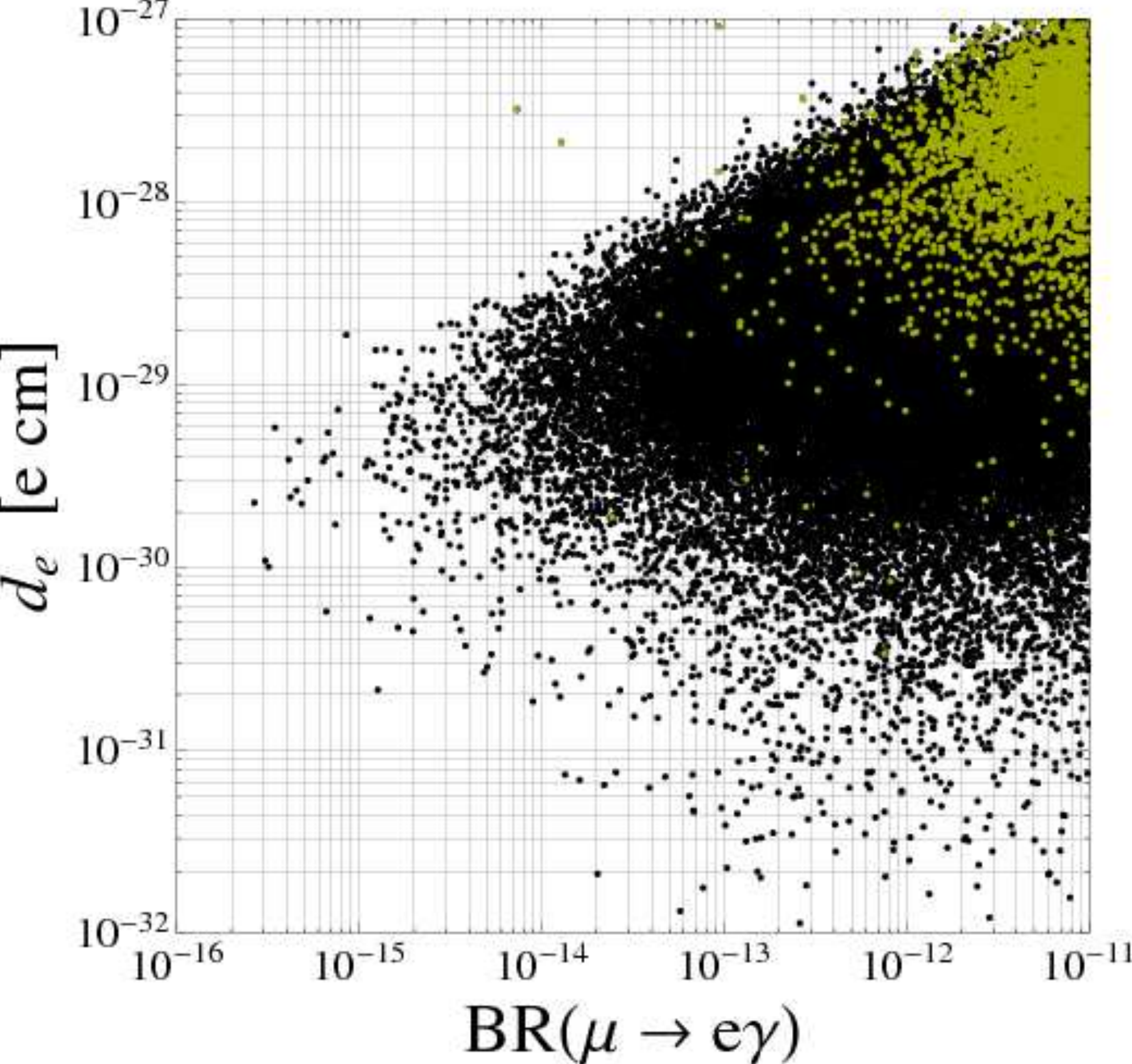}
\caption{\label{fig:buras}
$Br(\mu\to e\gamma)$ vs. $S_{\psi\phi}$ (left)
and $d_e$ vs. $Br(\mu\to e\gamma)$ (right) in the RVV model. 
The green points are consistent with the $(g-2)_\mu$
anomaly at $95\%$ C.L., i.e. $\Delta a_\mu\ge 1\times 10^{-9}$. 
From \cite{buras}.}
\end{center}
\end{figure}

As an example consider the supersymmetric non-Abelian flavour model of Ross,
Velasco and Vives (RVV), which leads to interesting correlatons between
quark- and lepton-flavour changing processes and also between CP-violation
in the quark and the lepton sector \cite{buras}. In the Standard Model
the mixing induced CP asymmetry in the $B_s$ system is predicted to be very 
small: $(S_{\psi\phi})_{\rm SM}\approx 0.04$. However, present data from
CDF and D0 could be the first hint for a  much larger value 
\cite{punzi,buras}
\begin{equation}
S_{\psi\phi}=0.81^{+0.12}_{-0.32}\ . 
\end{equation}
More precise measurements by CDF, D0, LHCb, ATLAS and CMS will clarify this 
intriguing puzzle in the coming years. In the RVV model, the prediction for
$S_{\psi\phi}$ is correlated with predictions for the branching ratio  
$Br(\mu\to e\gamma)$ and the electric dipole moment $d_e$ 
(see Fig.~\ref{fig:buras}). Consistency with the $(g-2)_\mu$ anomaly favours
smaller superparticle masses, which leads to a larger electric dipole moment 
and branching ratios $Br(\mu\to e\gamma)$ within the reach of the MEG
experiment at PSI \cite{bevan}. 

An important part of flavour physics is neutrino physics, currently an
experimentally-driven field \cite{wark}. The next goal is the measurement
of the mixing angle $\theta_{13}$ in the PMNS-matrix, with important
implications for the feasibility to observe CP violation in neutrino 
oscillations. Even more important is the determination of the absolute
neutrino mass scale. Cosmological observations have the potential to 
reach the sensitivity $\sum m_{\nu} < 0.1\ \mathrm{eV}$, which would be
very interesting for the connection to grand unification and also leptogenesis.
A model-independent mass determination is possible by measuring the endpoint
in Tritium $\beta$-decay where the KATRIN experiment is expected to 
reach a sensitivity of $0.2\ \mathrm{eV}$. 

\begin{figure}[b]
\begin{center}
\includegraphics[height=6.5cm]{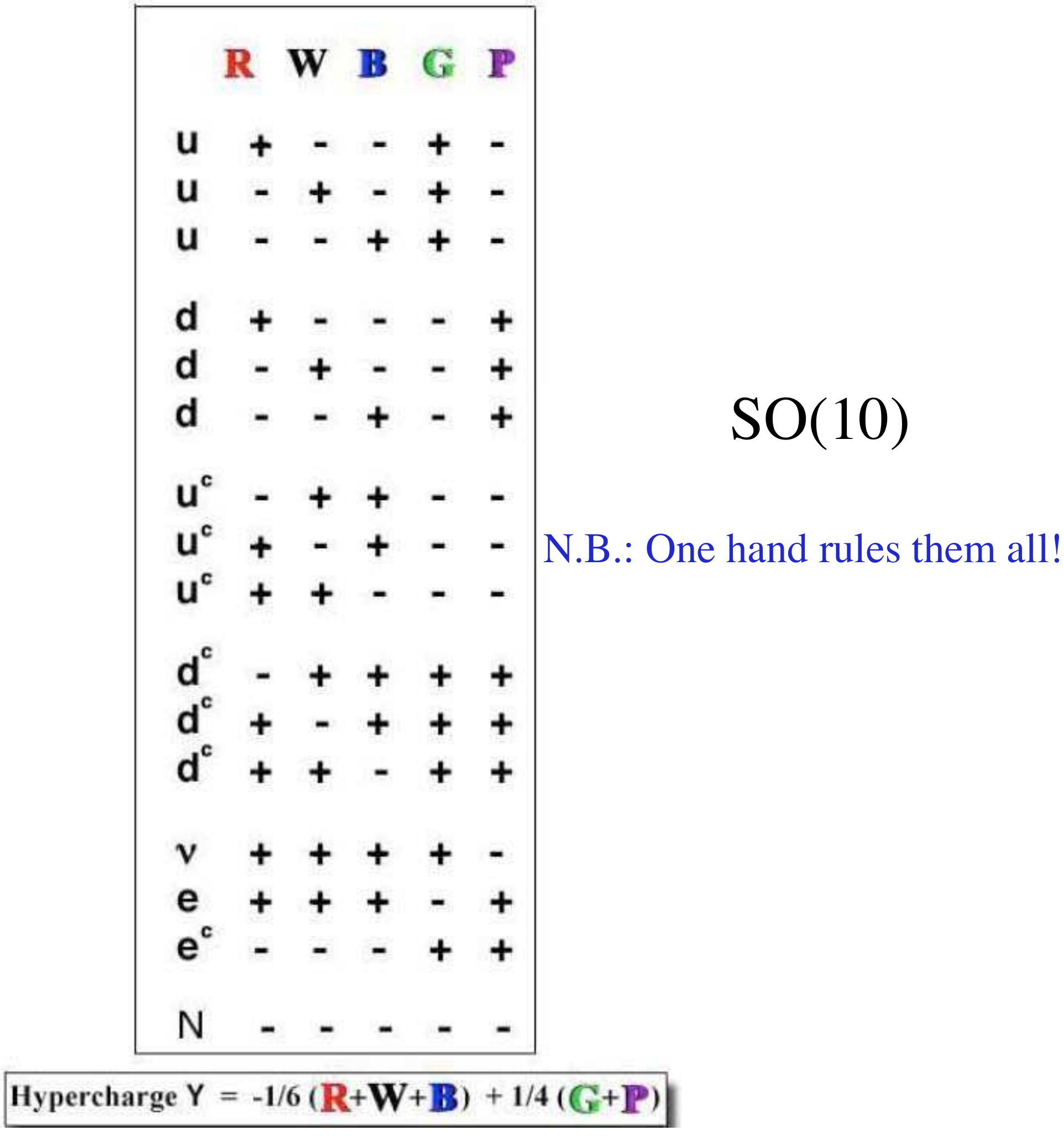}
\hspace*{1cm}
\includegraphics[width=7.5cm]{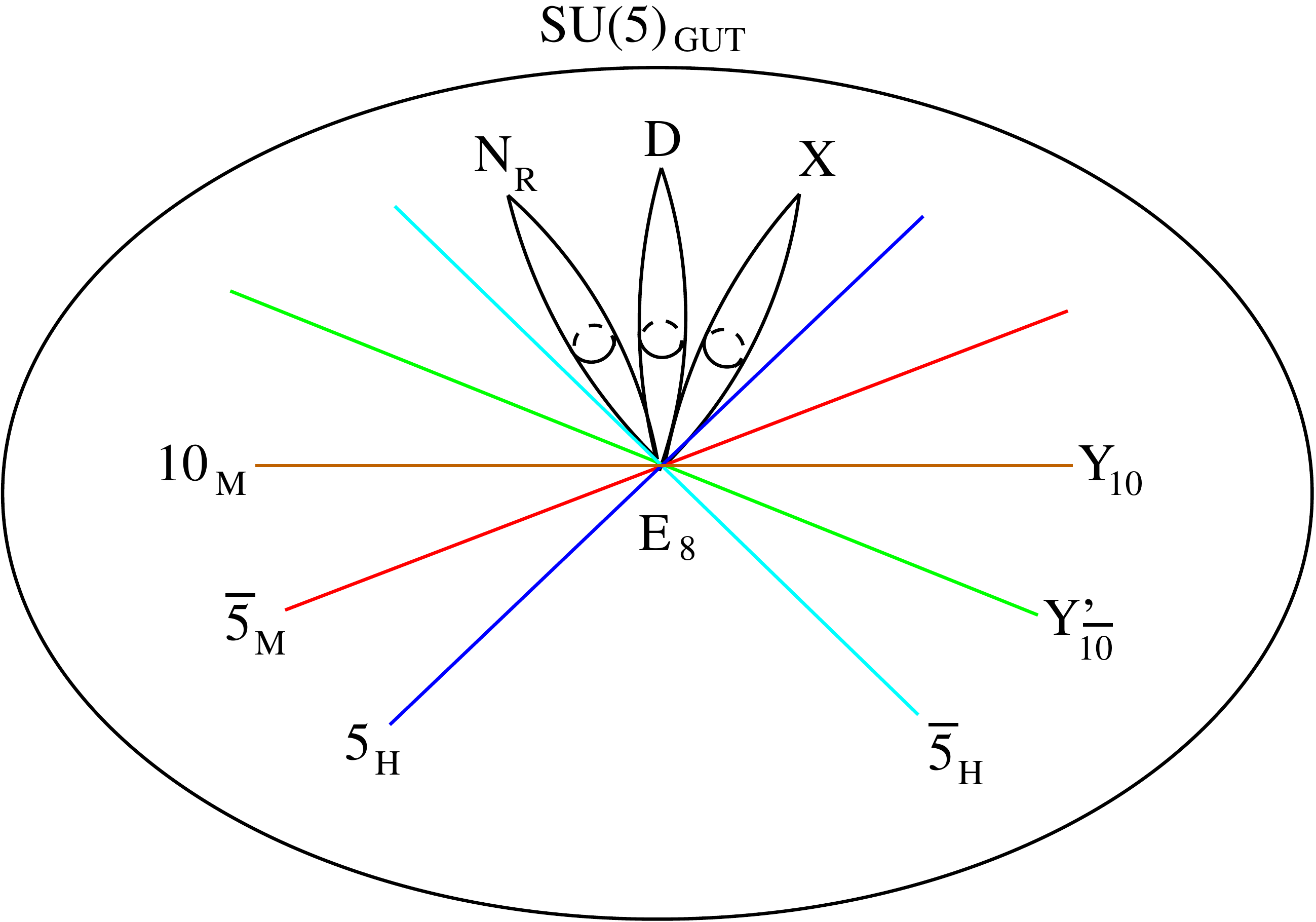}
\caption{ \label{fig:GUT}
Left: The unification group $SO(10)$ incorporates the Standard Model group
$SU(3)\times SU(2)\times U(1)$ as subgroup; the quarks and leptons of one
family, together with a right-handed neutrino, are united in a single
${\bf 16}$-plet of $SO(10)$; from \cite{wilczek}. Right: Geometric
picture of F-theory GUTs; matter and Higgs fields are confined to 
six-dimensional submanifolds; they intersect at a four-dimensional `point'
with enhanced $E_8$ symmetry where Yukawa couplings are generated.
From \cite{uranga}.  
}
\label{fig:wilczek}
\end{center}
\end{figure}

\section{GUTs and Strings}

The symmetries and the particle content of the Standard Model point towards
grand unified theories (GUTs) of the strong and electroweak interactions.
Assuming that the celebrated unification of gauge couplings in the 
supersymmetric Standard Model is not a misleading coincidence, supersymmetric
GUTs \cite{wilczek} have become the most popular extension of the Standard 
Model. Remarkably, one generation of matter, including the right-handed 
neutrino, forms a single spinor representation of $SO(10)$ 
(see Fig.~\ref{fig:wilczek}). It therefore appears natural to assume an
underlying $SO(10)$ structure of the theory. The route of unification
continues via exceptional groups, terminating at $E_8$,
\begin{equation}
SU(3)\times SU(2)\times U(1) \subset SU(5) \subset SO(10) \subset
E_6 \subset E_7 \subset E_8\ .
\end{equation}
The right-handed neutrino, whose existence is predicted by $SO(10)$ 
unification, leads to a successful phenomenology of neutrino masses and
mixings via the seesaw mechanism
and can also account for the cosmological matter-antimatter asymmetry via
leptogenesis.

The exceptional group $E_8$ is beautifully realized in the heterotic string.
Nonetheless, embedding the Standard Model into string theory has turned
out to be extremely difficult, possibly because of the huge number of string
vacua. Searching for the Standard Model vacuum in string theory would then
be like looking for a needle in a haystack. Recently, this situation has
improved, and promising string vacua have been found by incorporating 
GUT structures in specific string models \cite{uranga}. The different
constructions are based on Calabi-Yau or orbifold compactifications of
the heterotic string, magnetized brane models and, most recently, on F-theory
(see Fig.~\ref{fig:wilczek}). One obtains an appealing geometric picture
where gauge interactions are eight-dimensional, matter and Higgs fields are
confined to six dimensions, and Yukawa couplings are generated at the 
intersection of all these submanifolds, at a four-dimensional `point' with
enhanced $E_8$ symmetry.

The programme to embed the Standard Model into string theory using GUT
structures is promising but a number of severe problems remain to be solved. 
They include 
the appearance of states with exotic quantum numbers which have to be
removed from the low-energy theory, the treatment of supersymmetry breaking
in string theory and the stabilization of moduli fields. Optimistically, 
one can hope to
identify some features which are generic for string compactifications 
leading to the Standard Model, so that eventually string theory may lead
to predictions for observable quantities.

\section{Astrophysics and Cosmology}

\begin{figure}[b]
\includegraphics[width=6cm,angle=90]{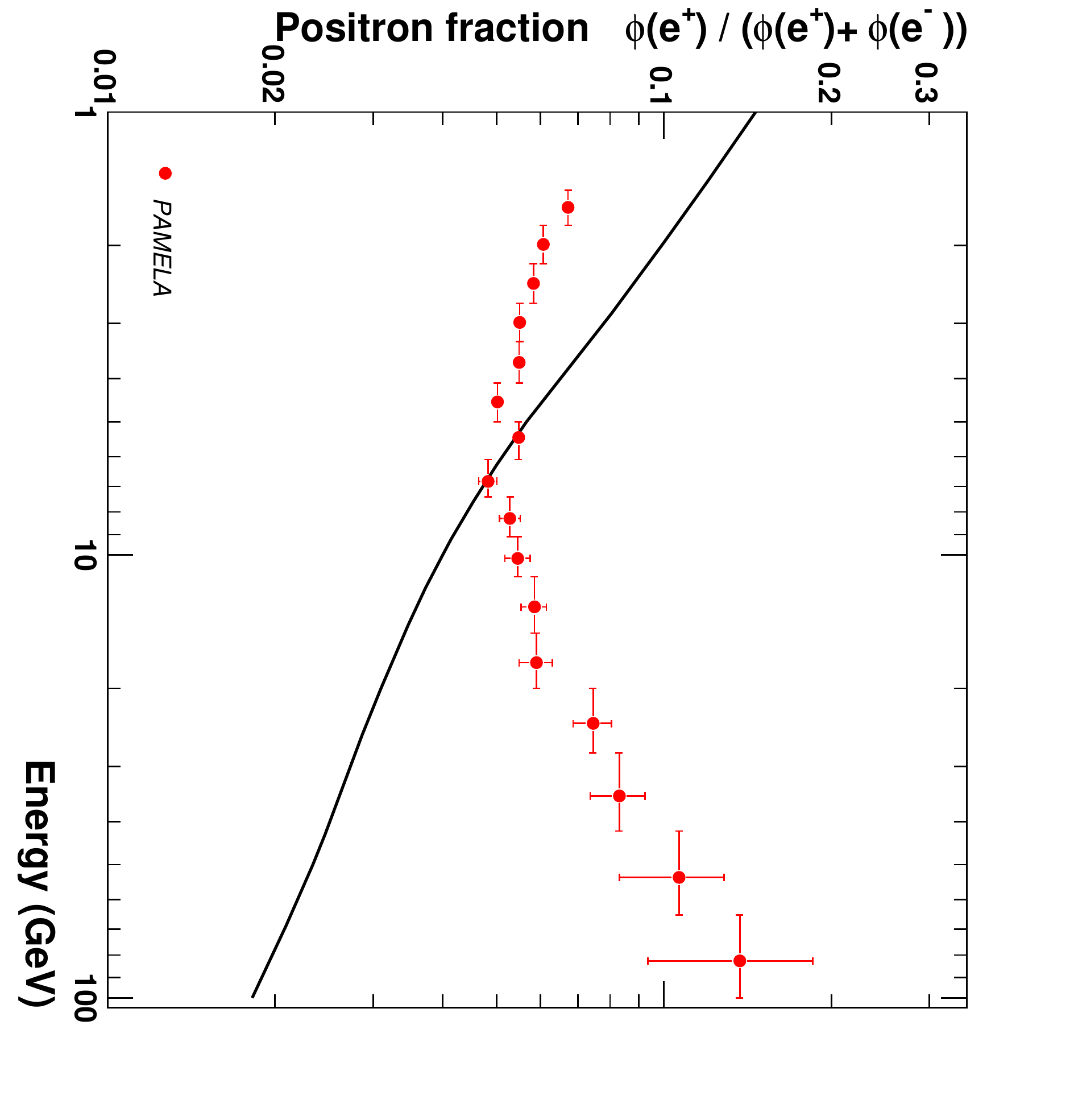}
\hspace*{0.6cm}
\includegraphics[width=8.1cm]{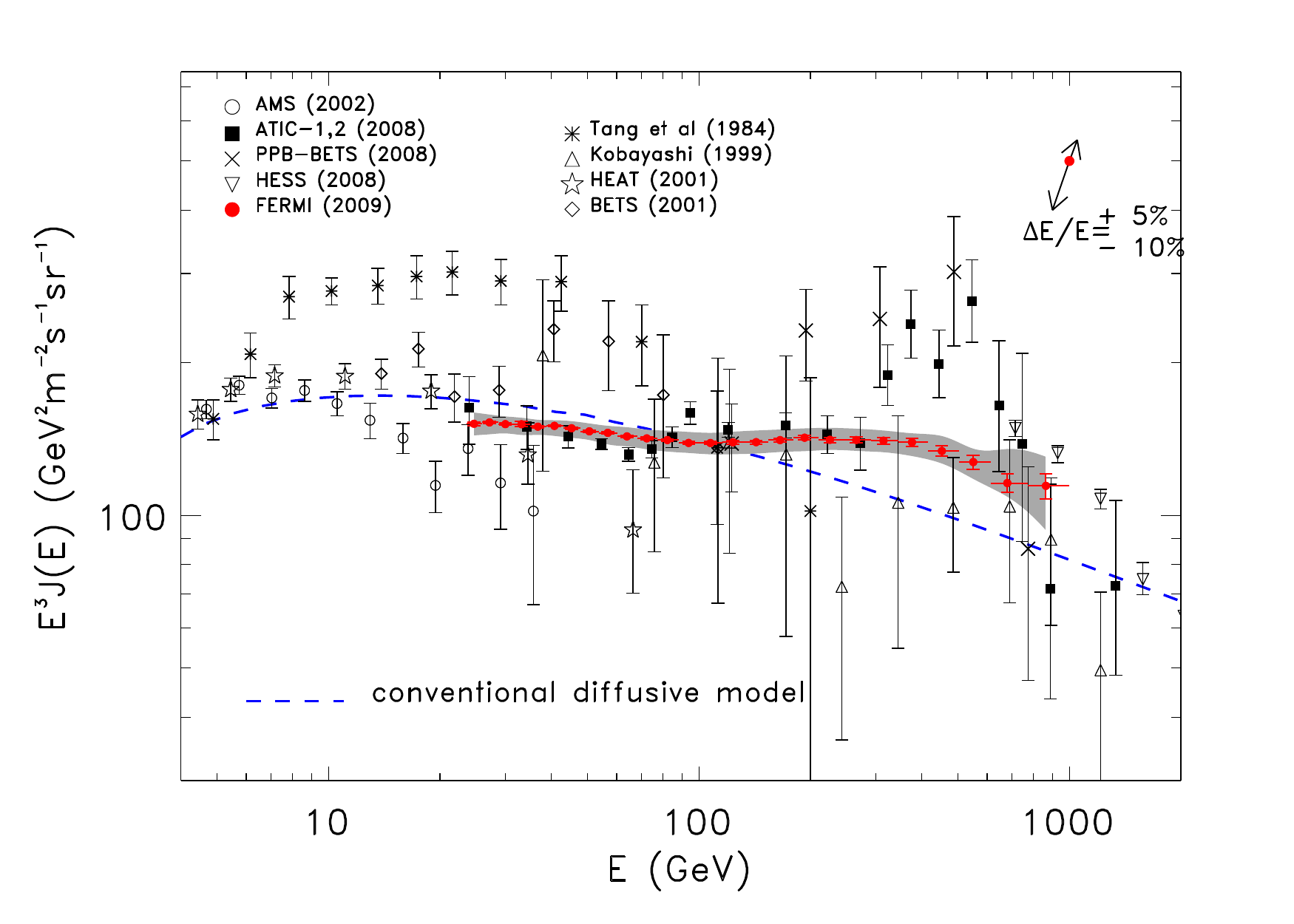}
\caption{
Left: The PAMELA positron fraction compared with the theoretical
model of Moskalenko \& Strong; the error bars correspond to one
standard deviation.
Right: The Fermi-LAT and HESS CR electron spectrum (red filled circles);
systematic errors are shown by the gray band; 
other high-energy measurements and a conventional diffusive model are also 
shown. From \cite{reimer}.
\label{fig:pamela}
}
\end{figure}

During the past year the cosmic-ray (CR) excesses observed by the PAMELA, 
Fermi-LAT and HESS collaborations (see Figs.~\ref{fig:pamela} and 
\ref{fig:strumiabest}) have received enormous attention \cite{reimer,strumia}.
This interest is due to the fact that the PAMELA positron fraction 
$e^+/(e^-+e^+)$ and the Fermi-LAT CR electron spectrum ($e^-+e^+$ flux) show
an excess above conventional astrophysical predictions at energies close to
the scale of electroweak symmetry breaking. This suggests that the observed
excesses may be related to dark matter consisting of WIMPs, Weakly Interacting
Massive Particles.

\begin{figure}[b]
\begin{center}
\hspace*{-1cm}
\includegraphics[width=1.05\textwidth]{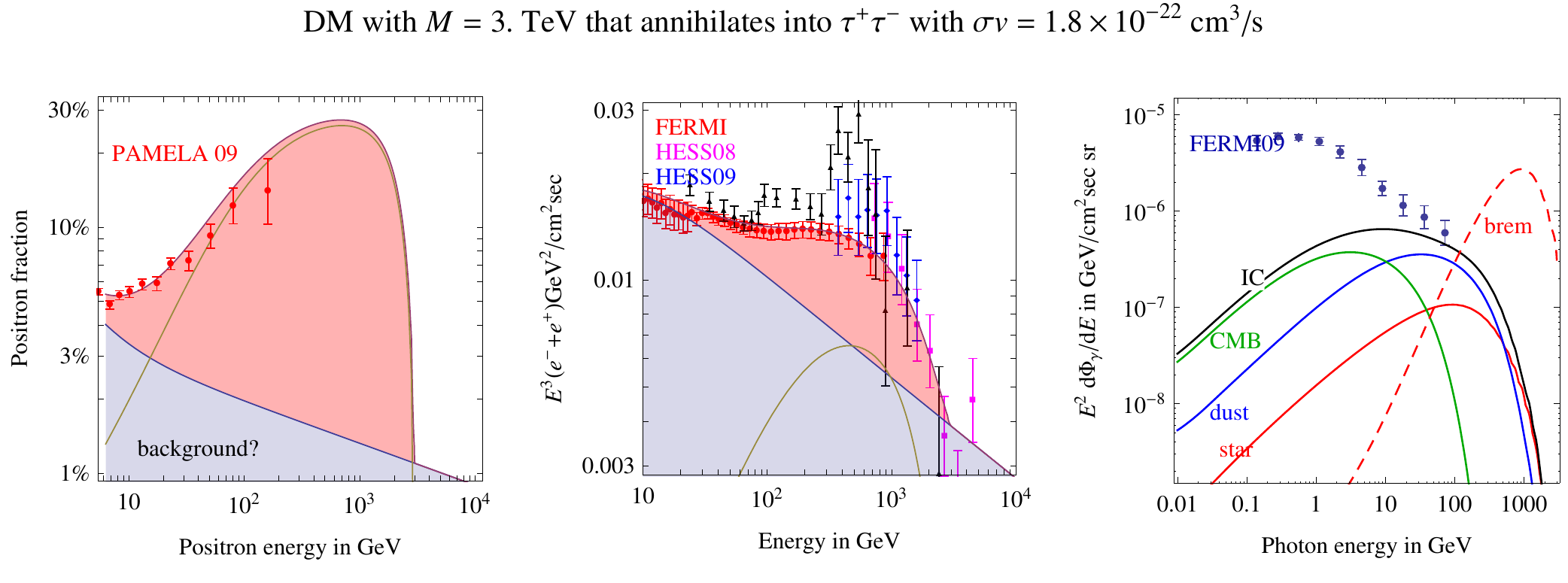}
\caption{
{\bf DM annihilations into $\tau^+\tau^-$}. The predictions are based on MED 
diffusion and the isothermal profile. Left: Positron fraction compared with
the PAMELA excess. Middle: $e^++e^-$ flux compared with the Fermi-LAT and
HESS data. Right: The Fermi-LAT diffuse gamma-spectrum compared with
bremsstrahlung (dashed red line) and inverse compton (IC) radiation (black
full line) with the components CMB (green), dust (blue) and CMB( green).
From \cite{strumia}.
\label{fig:strumiabest}}
\end{center}
\end{figure}
In the meantime various analyses have shown that both CR excesses can be 
accounted for by conventional astrophysical sources, in particular nearby
pulsars and/or supernovae remnants. On the other hand, it is still conceivable
that the excesses are completely, or at least partially, due to dark matter.
Since this is the main reason for the interest of a large community in
the new CR data, I shall focus on the dark matter interpretation in the 
following.

The first puzzle of the rising PAMELA positron fraction was the absence of an
excess in the antiproton flux. This led many theorists consider `leptophilic'
dark matter candidates where annihilations into leptons dominate over 
annihilations into quarks. The Fermi-LAT excess in $e^++e^-$ flux extends to 
energies up to a cutoff of almost 1~TeV, determined by HESS. Obviously, this
requires leptonic decays of heavy DM particles, with masses beyond the reach 
of LHC. A representative example of a successful fit is shown in 
Fig.~\ref{fig:strumiabest}. Note, that the gamma-ray flux due to 
bremsstrahlung and inverse Compton (IC) scattering of the produced leptons
is still consistent with present Fermi-LAT data. However, a remaining
problem of annihilating DM models is the explanation of the magnitude of
observed fluxes which is proportial $\langle \rho^2_{\mathrm{DM}} \rangle$,  
the square of the DM density. Typically, a large `boost factor', i.e.,
an enhancement of  $\langle \rho^2_{\mathrm{DM}} \rangle$ compared to
values obtained by numerical simulations, has to be assumed to achieve
consistency with observations. 

\begin{figure}[t]
\begin{center}
\hspace{-0.5cm}
\includegraphics[width=0.99\textwidth]{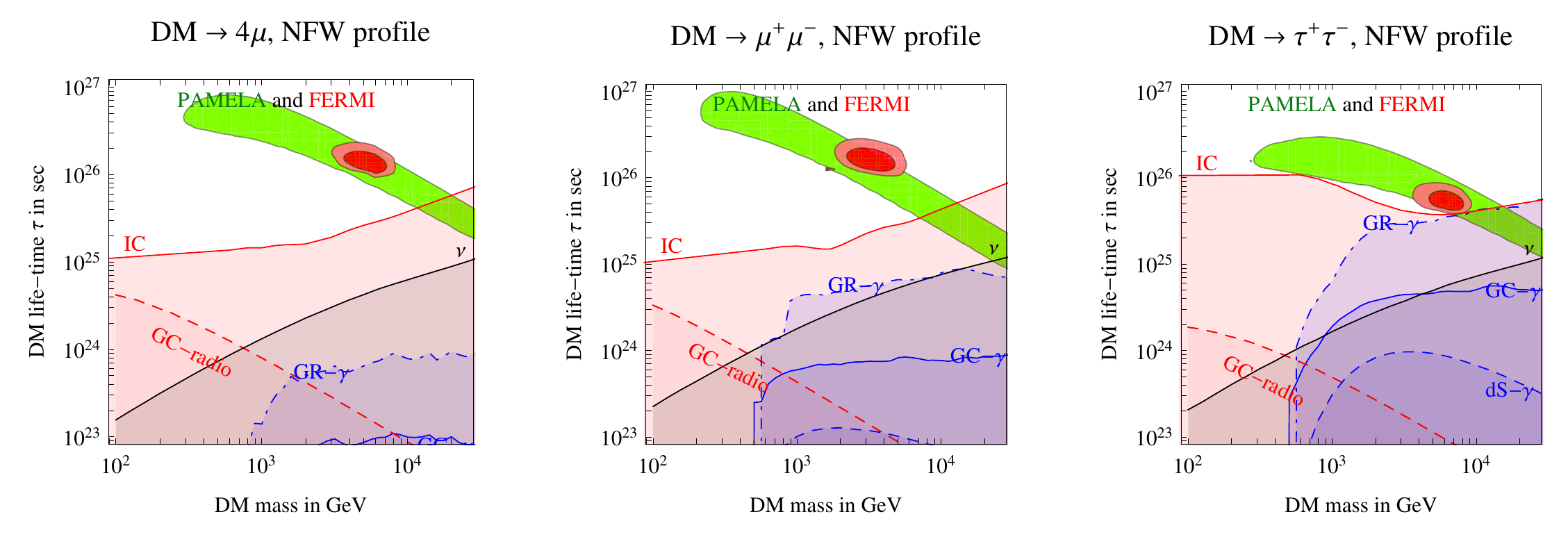}
\caption{{\bf DM decays into leptons}. Left: $4\mu$. Middle: $\mu^+\mu^-$.
Right: $\tau^+\tau^-$. Regions favored by PAMELA (green bands)
and by PAMELA, Ferimi-LAT and HESS observations (red ellipses) are
compared with HESS observations of the Galatic Center
(blue continuous line), the Galactic Ridge (blue dot-dashed)
and spherical dwarfes (blue dashed). From \cite{strumia}.
\label{fig:strumiadecay}.}
\end{center}
\end{figure}

The problems of annihilating DM models caused a new interest in decaying
DM models. Representative examples of dark matter candidates with different
leptonic decay channels are compared in Fig.~\ref{fig:strumiadecay}. Again
masses in the TeV range are favoured. The typical lifetime of $10^{26}$~s
is naturally obtained for decaying gravitinos, which can also be consistent
with the nonobservation of an antiproton excess, and in models where decays
are induced by GUT-suppressed dimension-6 operators. Decaying DM matter
models also lead to characteristic signatures at LHC, which are currently
actively investigated. 

\begin{figure}[b]
\includegraphics[width=5.5cm]{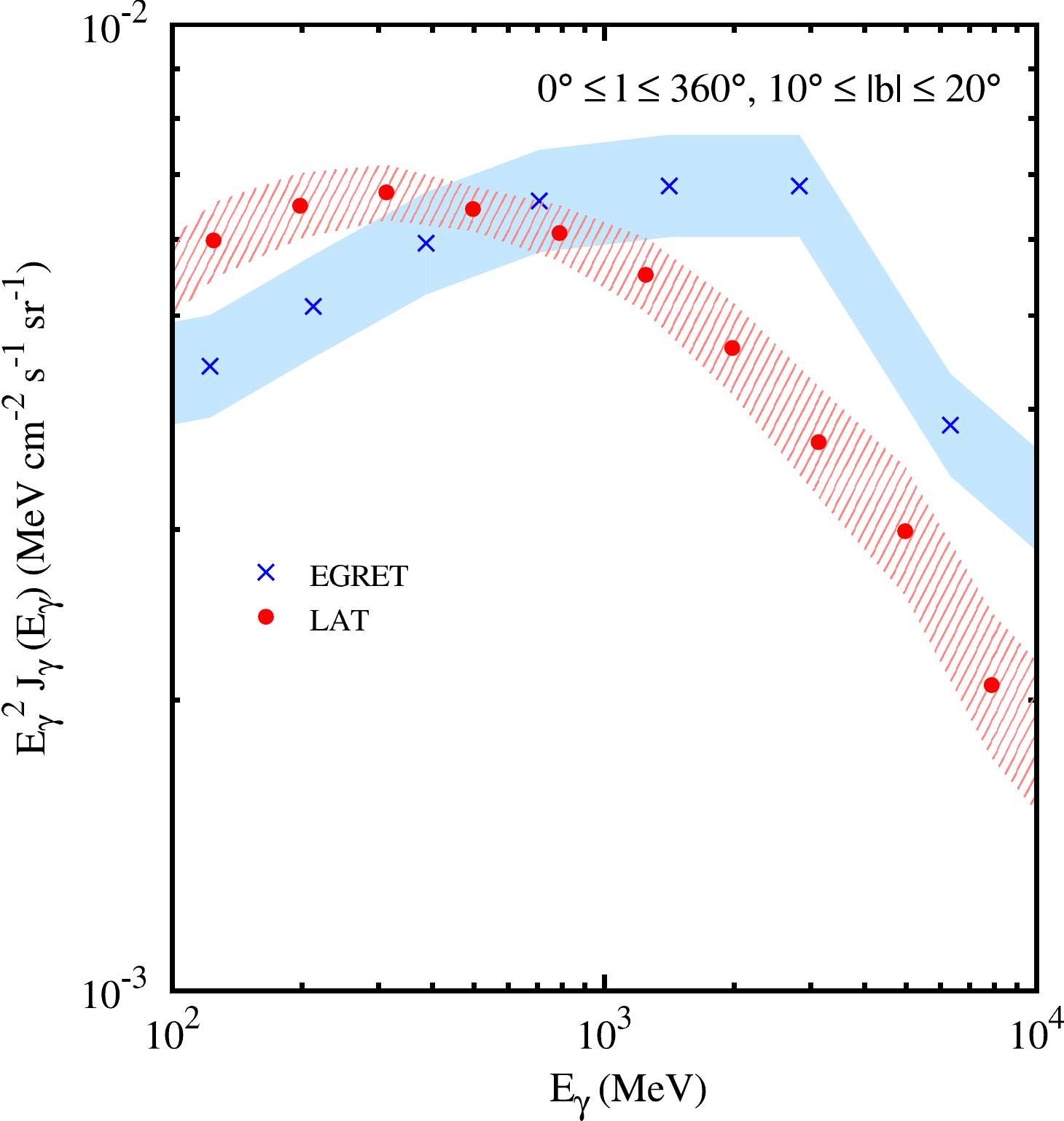}
\hspace*{1cm}
\includegraphics[width=8.1cm]{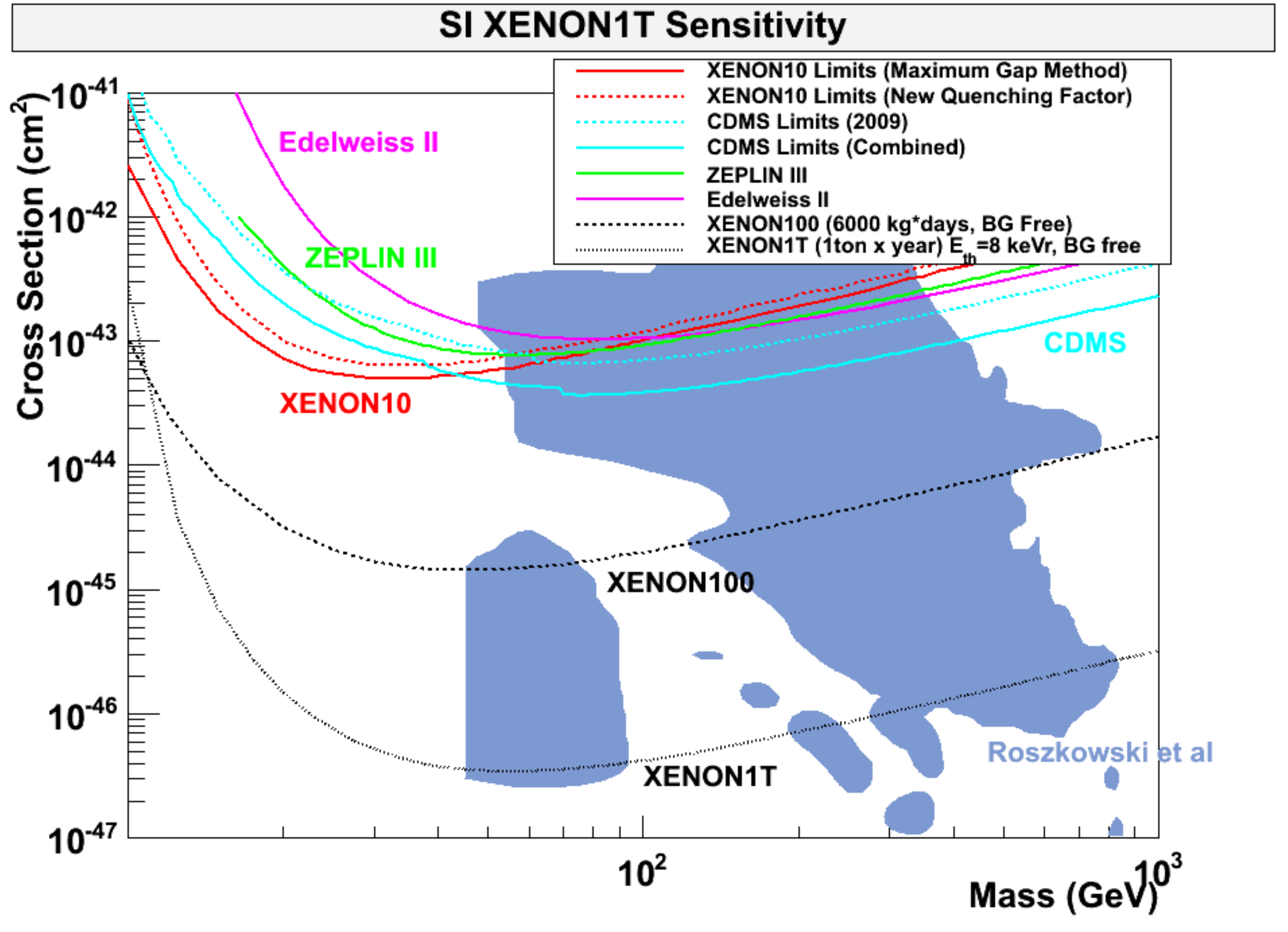}
\caption{Left: Galactic diffuse emission; intensity averaged over all
longitudes and latitudes in the range $10^\circ \leq |b| \leq 20^\circ$; 
data points and systematic uncertainties: Fermi-LAT (red), EGRET (blue);
from \cite{reimer}. Right: Present and projected bounds on spin-independent 
WIMP-nucleon cross sections from different experiments compared with 
predictions of Roszkowski et al. for supersymmetric models; from \cite{agile}.
\label{fig:diffuse}
}
\end{figure}

In addition to the $e^-+e^+$ flux the diffuse gamma-ray spectrum measured
by Fermi-LAT is of great importance indirect dark matter searches. Data for
the Galactic diffuse emission are shown in Fig.~\ref{fig:diffuse}. The
GeV excess observed more than 10 years ago by EGRET, which stimulated
several dark matter interpretations, has not been confirmed. Soon expected
data on the isotropic diffuse gamma-ray flux will severly constrain
decaying and annihilating dark matter models. In direct search experiments
limits on nucleon-WIMP cross sections have also been significantly improved.
The sensitivity will be further increased by two to four orders of magnitude 
in the
coming years (see Fig.~\ref{fig:diffuse}). They now probe a large part
of the parameter space of supersymmetric models, and in the next few years
we can expect stringent tests of WIMP dark matter from combined analyses of
direct and indirect searches, and LHC data. 
 
\begin{figure}[t]
\begin{center}
\includegraphics[width=9cm]{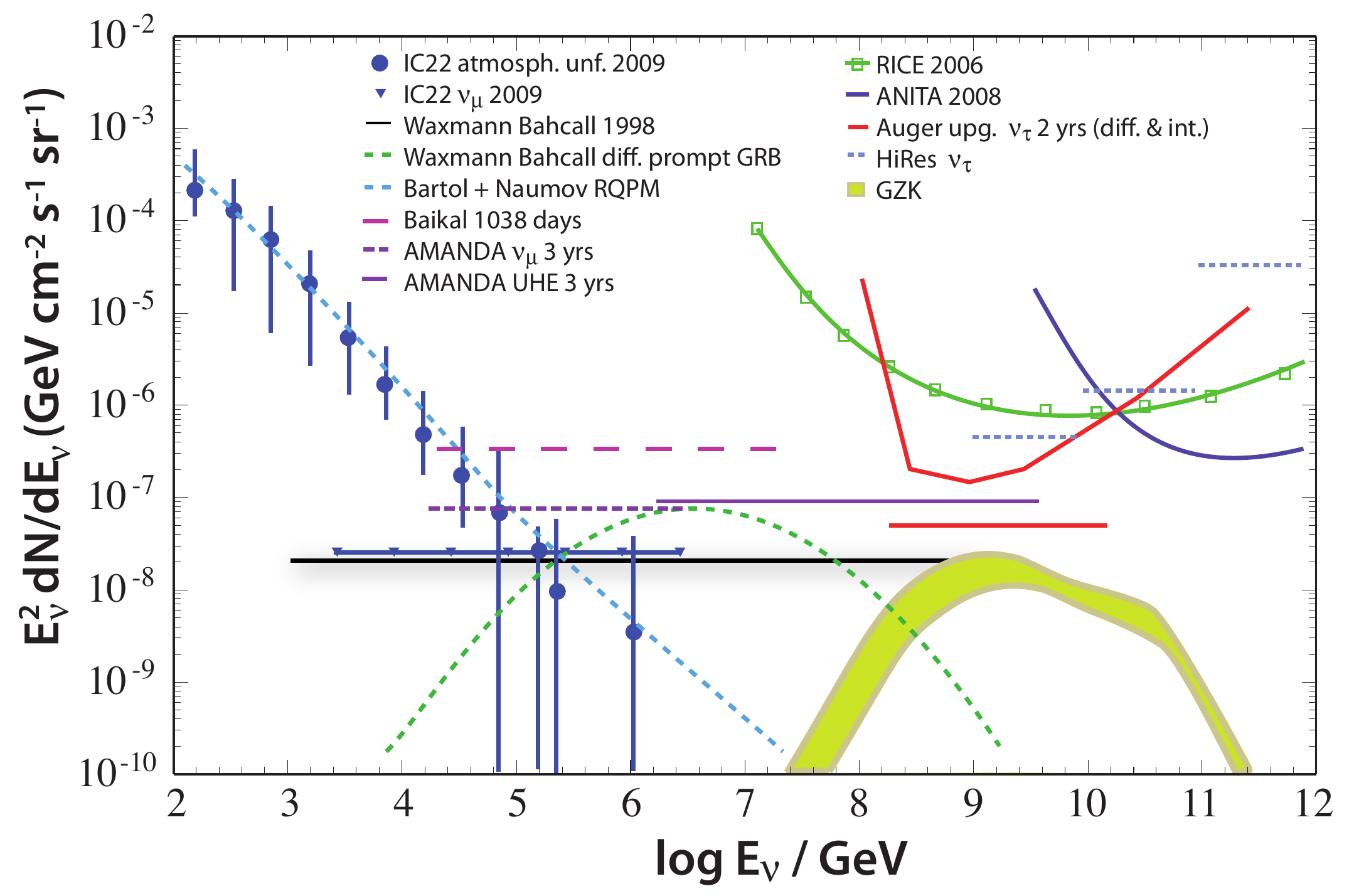}
\end{center}
\caption{Measured atmospheric neutrino fluxes and compilation of latest
limits on diffuse neutrino fluxes compared to predicted fluxes.
From \cite{kampert}.}
\label{fig:kampert}
\end{figure}

Annihilation of dark matter particles can also lead to high-energy neutrinos
which could be observed by large volume Cerenkov detectors such as AMANDA,
ANTARES and ICECUBE. Searches for diffuse neutrino fluxes have been performed
by a large number of experiments operating at different energy regions
(see Fig.~\ref{fig:kampert} for a compilation of recent data). The current
limits are approaching both the Waxmann-Bahcall and the cosmogenic flux
(labelled `GZK') predictions \cite{kampert}. 

Rapid advances in observational cosmology have led to a cosmological
Standard Model for which a large number of cosmological parameters have
been determined with remarkable precision. The theoretical framework is a 
spatially flat Friedman Universe with accelerating expansion \cite{mukhanov}.

\begin{figure}[t]
\begin{center}
\includegraphics[width=7.5cm]{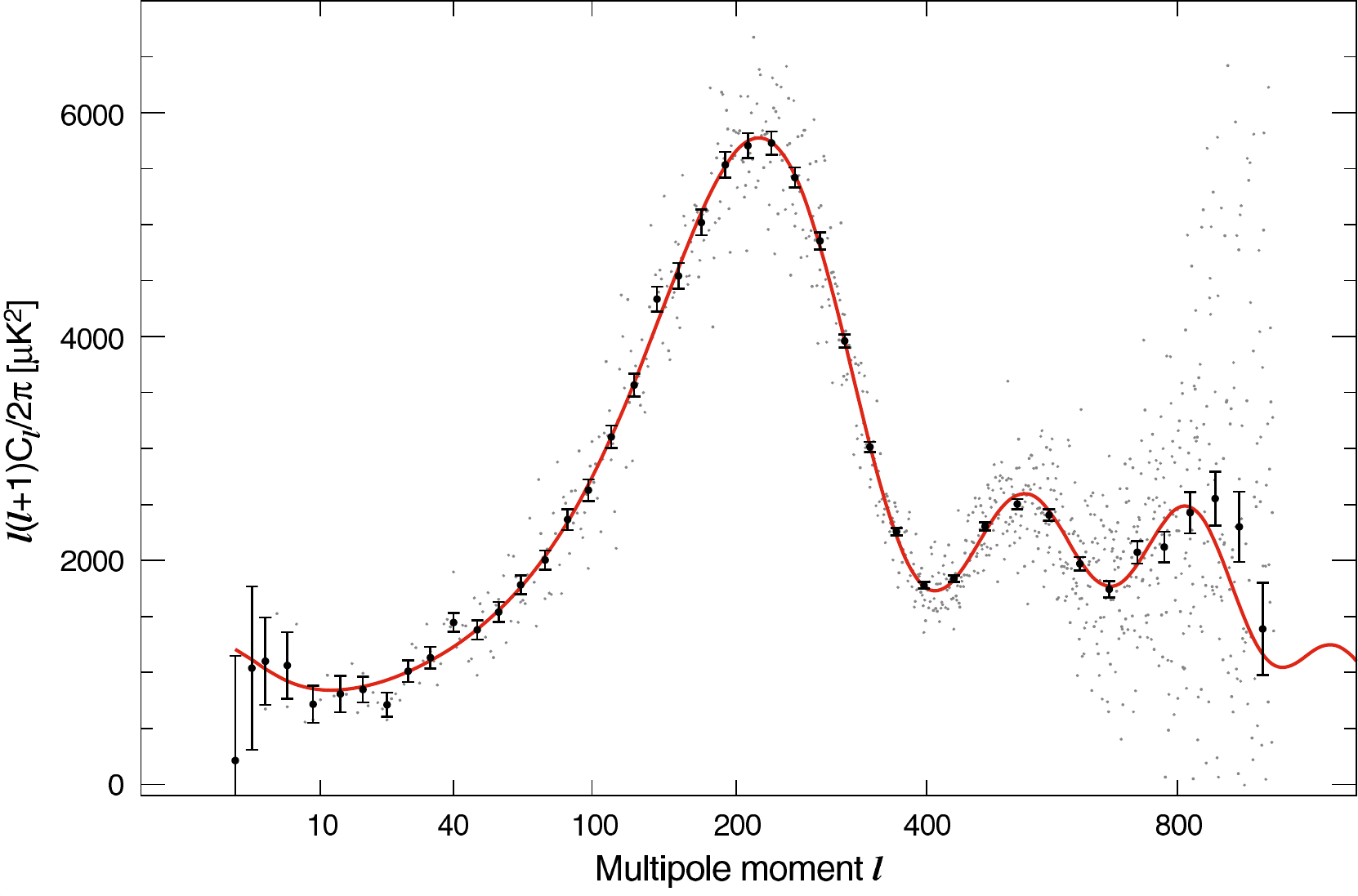}
\hspace*{1cm}
\includegraphics[height=6cm]{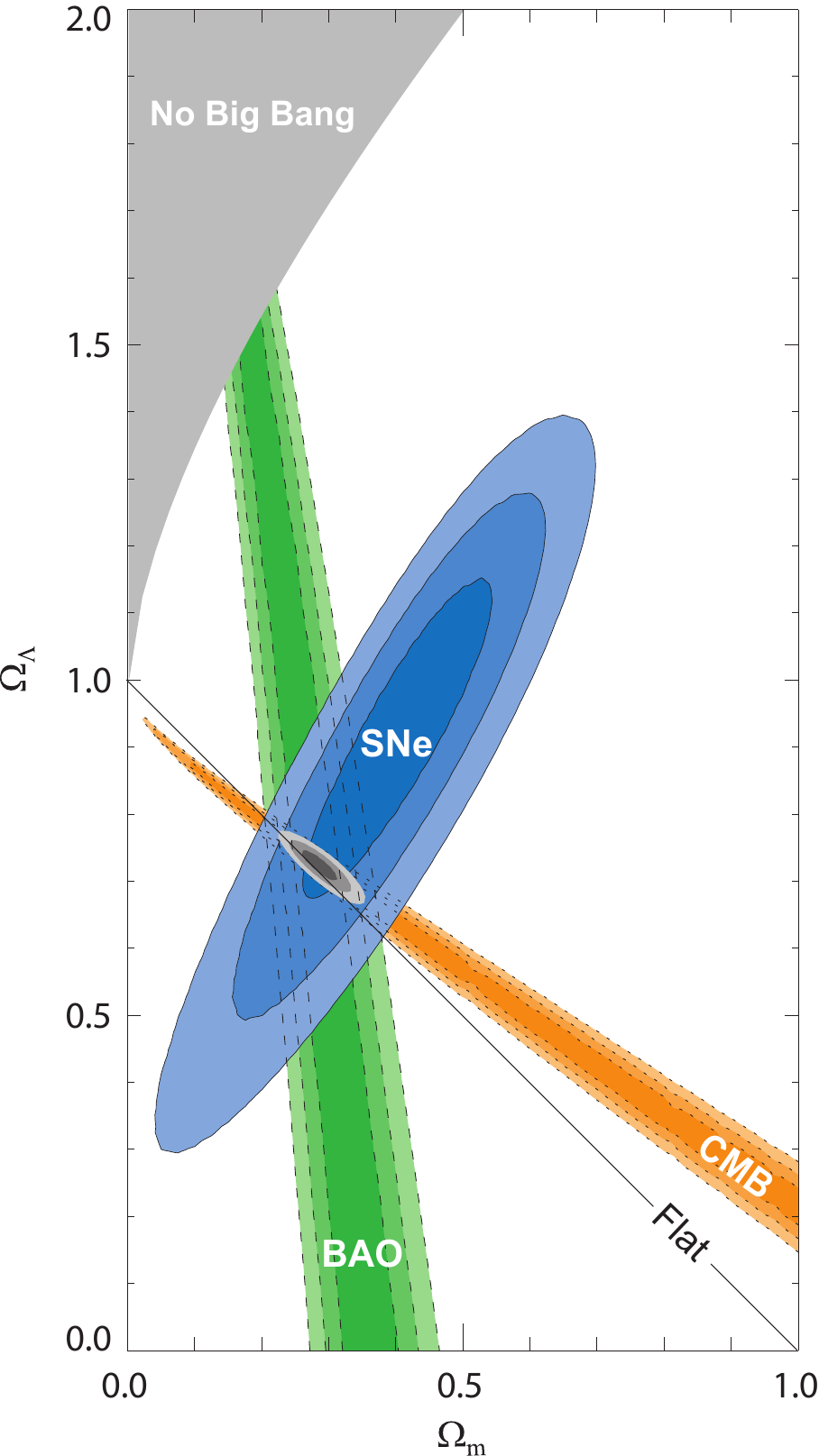}
\caption{ \label{fig:mukhanov}
Left: The angular power spectrum of the CMB temperature anisotropies
from WMAP5; the grey points are the unbinned data and the solid points
are binned data with error estimates; the solid line shows the prediction 
from the best fitting $\Lambda\mathrm{CDM}$ model.
Right: Confidence level contours of 68\%, 95\% and 99\% in the 
$\Omega_{\Lambda}-\Omega_m$ plane from the Cosmic Microwave Background, 
Baryonic Acoustic Oscillations and the Union SNe~Ia set, together with 
their combination assuming $w =-1$. From \cite{mukhanov}.  
}
\label{fig:mukhanov}
\end{center}
\end{figure}

The measurement of the luminosity distance of Type Ia supernovae (SNe~Ia),
used as `standard candles', and the analysis of the temperature anisotropies
of the cosmic microwave background (CMB) (see Fig.~\ref{fig:mukhanov}) have 
provided an accurate knowledge of the composition of the energy density of
the universe. This includes the total energy density $\Omega_{\mathrm{tot}}$,
the total matter density $\Omega_{\mathrm{m}}$, the baryon density 
$\Omega_{\mathrm{b}}$, the radiation density $\Omega_{\mathrm{r}}$, the 
neutrino density $\Omega_{\nu}$ and the cosmological constant 
$\Omega_{\Lambda}$; the total matter density contains the
cold dark matter density, 
$\Omega_{\mathrm{m}} = \Omega_{\mathrm{cdm}} + \Omega_{\mathrm{b}}$.
Most remarkably, the universe is spatially flat within errors,
\begin{equation}
\Omega_{\mathrm{tot}} = 1.006 \pm 0.006\ ,
\end{equation}
and dominated by dark matter ($\Omega_{\mathrm{cdm}} \simeq 0.22$) and
dark energy ($\Omega_{\Lambda} \simeq 0.74$) \cite{mukhanov}.

In the future, gravitational waves may become a new window to the present
as well as the early universe. Impressive progress has been made in 
improving the sensitivity of current laser interferometers, and detection
of gravitational waves is definitely expected with the next generation of
detectors \cite{danzmann}. If the sensitivity can be increased to higher
frequences, it is conceivable that the equation of state in the very
early universe can be probed with gravitational waves.

\begin{figure}[t]
\begin{center}
\includegraphics[width=6cm]{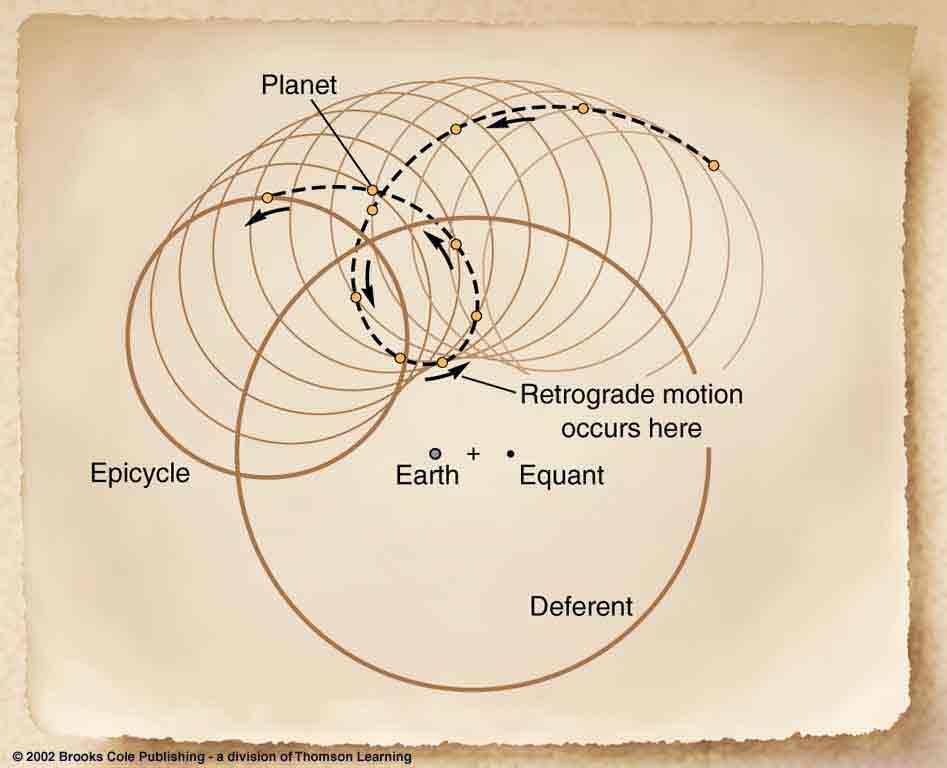}
\vspace*{1cm}\hspace*{1cm}
\includegraphics[width=6cm]{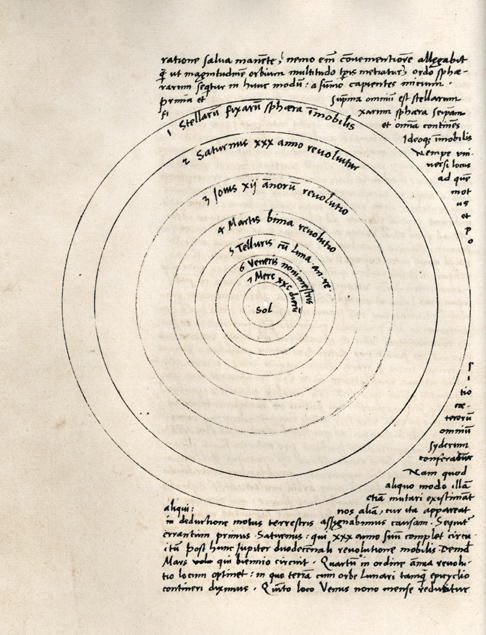}
\caption[]{Left: Ptolemy's epicycle model of the planetary system. Right: 
Copernicus's heliocentric model of the planetary system.}
\label{fig:epicycle}
\end{center}
\end{figure}

Many dark matter candidates have been suggested in various extensions of
the Standard Model of Particle Physics and we can hope that new data form
the LHC, and direct and indirect search experiments will clarify this
problem in the coming years. On the contrary, research on `dark energy' 
is dominated by question marks \cite{mukhanov}. Many explanations have been
suggested, including quintessence, k-essence, modifications of gravity,
extra dimensions etc., but experimental and theoretical breakthroughs still
appear to be ahead of us.  

\section{Outlook}

With the start of the LHC \cite{evans} and new data taken by ATLAS 
\cite{gianotti}, CMS\cite{virdee}, LHCb \cite{golutvin} and ALICE
\cite{giubellino} we are entering a new era in Particle Physics. We expect
to gain deeper insight into the mechanism of electroweak symmetry breaking,
the origin of quark and lepton mass matrices and the behaviour of matter
at high temperatures and densities,
and many hope that supersymmetry will be discovered. 

Important results can also be expected from ongoing and planned 
non-accelerator experiments, cosmic-ray experiments and more precise
cosmological observations. These include the determination of the absolute 
neutrino mass scale, possible evidence for weakly interacting dark
matter particles, polarization of the cosmic microwave background
and the determination of the equation of state of dark energy for
different redshifts.
  
On the theoretical side, there appear to be two main avenues beyond the
Standard Model: (A) New strong interactions at TeV energies, like composite
W-bosons, a composite top-quark, technicolour or large extra dimensions,
or (B) the extrapolation of the Standard Model far beyond the electroweak
mass scale, with more and more symmetries becoming manifest: supersymmetry,
grand unified symmetries, higher-dimensional space-time symmetries and
possibly symmetries special to string theory.

In Cracow we are reminded of Nicolaus Copernicus who, about 500 years ago,
invented the heliocentric model of the planetary system, in contrast to 
Ptolemy's epicycle model (see Fig.~\ref{fig:epicycle}). Given the high
symmetry and simplicity of the heliocentric model, one may 
think that Copernicus would have had a preference for avenue (B) beyond the
Standard Model, but we obviously cannot be sure. It took about seventy years
until, after new astronomical observations, the heliocentric model was
generally accepted. Fortunately, with the successful start of the LHC, we can
hope for crucial information about the Physics beyond the Standard Model 
much faster.

\section*{Acknowledgements}

I would like to thank the members of the international and local organizing
committees, especially Antoni Szczurek and Marek Je\.zabek, for their
successful work and for their hospitality in this beautiful city. I am indepted
to many colleagues at this conference and at DESY for their help in the 
preparation of this talk, especially Andrzej Buras, Laura Covi, Leszek Motyka,
Peter Schleper and Fabio Zwirner.

\end{document}